\documentclass[pra,preprint,showpacs]{revtex4}
\usepackage[centertags]{amsmath}
\usepackage{amsfonts}
\usepackage{amssymb}
\usepackage{amsthm} 
\usepackage{newlfont}
\usepackage{epsfig}
\usepackage{amscd}
\usepackage{graphicx}
\usepackage{epsfig}
\usepackage{footnote}
\usepackage{lipsum}
\usepackage{color}
\usepackage{xcolor}
\usepackage{graphicx}
\usepackage{subfig}
\usepackage{amscd}

\newcommand{\beq}{\begin{equation}}
\newcommand{\eeq}{\end{equation}}
\newcommand{\ba}{\begin{array}}
\newcommand{\ea}{\end{array}}
\newcommand{\bea}{\begin{eqnarray}}
\newcommand{\eea}{\end{eqnarray}}
\begin{document}

\begin{center}
{\large \sc \bf {Relay entanglement and clusters of correlated spins. }
}

\vskip 15pt

{\large 
S.I.Doronin and A.I.~Zenchuk 
}

\vskip 8pt

{\it $^2$Institute of Problems of Chemical Physics, RAS,
Chernogolovka, Moscow reg., 142432, Russia}.

\end{center}


\begin{abstract}
Considering a spin-1/2 chain, we  {suppose} that the entanglement passes from a given pair of particles to another one,  thus establishing the relay transfer of entanglement along  the chain. Therefore, we introduce the relay entanglement as a sum of all pairwise entanglements in a spin chain. For more  detailed studying the effects of {remote} pairwise entanglements, we use  the partial sums collecting entanglements between the spins separated by up to a certain number of nodes.  The problem of entangled cluster formation is considered, and the  geometric mean  entanglement is introduced as a {characteristics} of quantum correlations in a cluster. Generally,  the life-time of a cluster decreases with an increase in its size.
\end{abstract}

\maketitle

\section{Introduction}
\label{Section:Introduction}

The quantum correlations are responsible for  a principal advantages of  quantum devises over their classical counterparts. 
The problems of correlations between remote systems \cite{BBVB,BZ}, formation  \cite{DSC,LBAW,NLLZ,ZC,SXSZDWHCKW,RDL} and { transfer of quantum correlations \cite{LS,SO,BACVV,HZI,DZ_2016}} are getting an increasing meaning because of progress in quantum information processing which includes quantum state creation and information transfer 
\cite{PBGWK,PBGWK2,BDSSBW,BHLSW,G}. Obviously, the  relation between the information transfer and quantum correlations exist; however it can  not be simply
released. Thus, the sender-receiver correlations (entanglement) were studied in set of papers \cite{Bose,VGIZ,DZ_2010,GMT,BB,DFZ_2009}. Nevertheless, many issues require clarification. For instance,  it was shown  that the {sender-receiver entanglement ($SR$-entanglement)  is vanishing in the case} of  perfect state transfer \cite{GMT} and { it also can be vanishing}  in the case of remote state creation \cite{DZ_2017}. It was shown in \cite{DZ_2017} that the possibility to transfer the information  {(i.e., the parameters of the sender's initial state)} from the sender to the  receiver is directly related to a certain determinant condition (associated { with} the informational correlation in 
Ref.\cite{ZInf_2014}). But still this determinant condition is not {straightforwardly} related to the traditional measures such as entanglement \cite{Wootters,HW} and discord \cite{HV,OZ,Z}.

{ It follows from the above  that quantum information transfer can be established without the $SR$-entanglement, and therefore, this entanglement can not be responsible for such a process \cite{GMT,DZ_2017}.  We emphasize that this fact  makes principal difference between the information transfer and teleportation \cite{BBCJPW,BPMEWZ,BBMHP} where the entanglement between the sender and receiver is necessary. 

Our paper is aimed at studying quantum correlations which can govern the state (information) transfer from the sender to the receiver. We assume that, similar to the simple explanation of the state transfer as an excitation transfer  from one node to another,  the 
mechanism of  entanglement transfer is basically the same. Namely, 
if at some instant  there are  two large pairwise entanglements, then we can say that the  entanglement passes between these pairs.} 
Therewith these pairs     can be the pairs of remote spins (not of the close neighbors only). { Thus, the entanglement is transferred along the chain during some evolution period $T$ if there are significant time-overlaps between the pairwise entanglements such that at each  instant $t$, $0<t\le T$,  there are at least two large pairwise entanglements. If this is the case, then 
 the sum of all pairwise entanglements is non-vanishing over the whole evolution   period $T$, although each individual pairwise entanglement is large only over the relatively small sub-interval of the  period $T$. 
 We propose that this propagating entanglement supplements the process of information propagation.  In other words, there is  an 
assembly of pairwise  entanglements (rather then the single $SR$-entanglement)   that governs the  information transfer along the spin chain.
This assembly is characterized by the above  sum which we refer to as  the  relay entanglement.  First this concept was introduced  in   \cite{FZ_2013}. Thus, the relay entanglement is a pairwise mechanism of entanglement propagation along the spin chain and  establishes  the   indirect $SR$-entanglement.  We emphasize that the indirect $SR$-entanglement found here does not assume arising the $SR$-entanglement at any instant of the evolution period $T$. In particular, the $SR$-entanglement can be absent during the whole evolution period $T$ as was observed in \cite{DZ_2017}, such behavior does not contradict our protocol. Another feature of the relay entanglement 
is that  it arises  during the evolution even if  there is no initial entanglement, unlike the  transfer of  entangled states from the multi-qubit sender to the  multi-qubit receiver \cite{LS,SO,BACVV} where  entanglement is present initially. 

Of course, the contribution to the relay entanglement from the remote pairs and from the close ones is different. For detailed study of these contributions, we introduce the so-called partial sums collecting the entanglements between the spins separated by up to a certain number of nodes.   We study the dependence of the partial sums  from the distance  between the correlated spins and show that the result significantly depends on the parameters of the sender's and receiver's initial states. In particular, the contribution from the remote spins can be more important in certain cases. }

{
In addition, in the course of evolution, the pairwise entanglements can form clusters, i.e., families of entanglements with large time-overlaps between each two of them. Accordingly, the set of neighboring spins, such that all possible pairwise entanglements in this set are significant at some  instant,  form the cluster of correlated spins at this  instant  \cite{THSE,KS,FZ_2013}. Each individual cluster exists during some time interval, then it   diffuses and another cluster appears.  
 We study the  cluster formation  in dependence on  the initial parameters of the sender and receiver states and estimate the life-time of a cluster.  
 As { a} characteristic of the entanglement in a cluster, we introduce a geometric mean of all the pairwise entanglements inside of this cluster.
 Formation of  entangled clusters in an evolutionary chain
{ represents} a particular way of creating the quantum subsystems with all entangled nodes. Such clusters can be  candidates for quantum registers of small size.}

The structure of the paper is following. The model of communication line we deal with is described in Sec.\ref{Section:model}. The relay entanglement in a symmetrical model as a characteristics of  entanglement propagation is studied in Sec.\ref{Section:relay}. The clusters of entangled particles arising during evolution are described in Sec.\ref{Section:clusters}. The obtained results are discussed  in Sec.\ref{Section:conclusion}.

\section{Model}
\label{Section:model}
We consider {a} communication line based on a homogeneous spin chain which includes the one-qubit  sender (S), transmission line (TL) and one-qubit receiver (R) { and 
evolves} under the 
nearest-neighbor $XX$-Hamiltonian {\cite{Mattis}}
\begin{eqnarray}
H=\sum_{i=1}^{N-1} D (I_{ix} I_{(i+1)x} + I_{iy} I_{(i+1)y}) .
\end{eqnarray}
The initial density matrix is a tensor product state:
\begin{eqnarray}
\rho_0=\rho^S_0\otimes \rho^{TL}_0 \otimes \rho^R_0,
\end{eqnarray}
where the transmission line $TL$ is in the ground state initially
\begin{eqnarray}
\rho^{TL} = |0\rangle_{TL}   ~_{TL}\langle 0|,
\end{eqnarray}
and the initial states of the 
sender $S$  and receiver $R$ are represented  as follows \cite{DZ_2016}:
\begin{eqnarray}\label{inSR}
\rho^S_0 = U^S \Lambda^S (U^S)^+,\;\; \rho^R_0 = U^R \Lambda^R (U^R)^+.
\end{eqnarray}
Here the eigenvalue and eigenvector matrices read, respectively,
\begin{eqnarray}\label{ev}
\Lambda^S={\mbox{diag}}(\lambda^S,1-\lambda^S),\;\;\;\Lambda^R={\mbox{diag}}(\lambda^R,1-\lambda^R),
\end{eqnarray}
and
\begin{eqnarray}\label{US}
U^S=\left(
\begin{array}{cc}
\cos \frac{\pi \alpha_1}{2} & - e^{-2 i \pi \alpha_2} \sin \frac{\pi \alpha_1}{2}\cr
e^{2 i \pi \alpha_2} \sin \frac{\pi \alpha_1}{2} & \cos \frac{\pi \alpha_1}{2} 
\end{array}
\right),
\end{eqnarray}
\begin{eqnarray}\label{UR}
U^R=\left(
\begin{array}{cc}
\cos \frac{\pi \beta_1}{2} & - e^{-2 i \pi \beta_2} \sin \frac{\pi \beta_1}{2}\cr
e^{2 i \pi \beta_2} \sin \frac{\pi \beta_1}{2} & \cos \frac{\pi \beta_1}{2} 
\end{array}
\right).
\end{eqnarray}
The parameters $\lambda^S$, $\lambda^R$ are referred to as the  eigenvalue control parameters, while 
the parameters  $\alpha_i$, $\beta_i$ ($i=1,2$) are called the
eigenvector control parameters. The term ''control'' means the responsibility of these parameters for the value of  entanglement.
The variation intervals for these parameters  are following:
\begin{eqnarray}
\label{intervals}
\begin{array}{ll}
0\le \alpha_i \le 1,& 0\le \beta_i \le 1,\;i=1,2,\cr
0\le \lambda^S \le 1, & 0\le \lambda^R \le 1.
\end{array}
\end{eqnarray}
Studying the evolution of pairwise correlations, we  need the density 
matrix of the subsystem of the $i$th and $j$th spins, which reads 
\begin{eqnarray}\label{RhoSR}
\rho^{(ij)}(t) ={\mbox{Tr}}_{/ij} \Big(V(t) \rho_0 V^+(t) \Big),
\end{eqnarray}
where the trace is taken over all the spins except the $i$th and $j$th ones.
In particular, if $i=1$ and $j=N$, we have {the density matrix for} the  $SR$-state.

The dynamics of communucation line with { the} one-qubit sender and receiver and ground initial state of TL reduces to the two-excitation subspace of the whole $2^N$-dimensional Hilbert space related to the $N$-node spin-1/2 system. This significantly simplifies calculations. Nevertheless, to provide a detailed analysis given below we restrict ourselves to the  spin chain of $N=10$ nodes.    

We consider the evolution of quantum correlations during the time interval from zero to  some optimized  instant. 
Following Ref.\cite{DZ_2017}, we chose 
 { such the instant}  {that the sum of all but one diagonal elements of the $SR$-density matrix $\rho^{(1N)}$ 
\begin{eqnarray}\label{s}
s=\sum_{i=2}^4 \rho^{1N}_{ii}
\end{eqnarray}
(we use the basis $|00\rangle$,  $|01\rangle$,  $|10\rangle$,  $|11\rangle$)
averaged over the parameters of the initial state
takes the maximal value.
 The  instant found in this way corresponds to the case when the maximal initial signal is collected at the sender and receiver, i.e., the  instant suitable for the state registration at the subsystem $SR$. } The choice of this  instant is predicted by our purpose of studying those correlations which can be responsible for the information transfer between the sender and receiver. 

We consider entanglement as a measure of quantum  correlations. The Wootters criterion \cite{Wootters,HW} allows us to  calculate the quantum 
entanglement between the nodes $n$ and $m$.  
Traditionally the entanglement $E$ can be represented as a  monotonic function of so-called 
concurrence $C$, $E=-\frac{1+\sqrt{1-C^2}}{2}\log_2 \frac{1+\sqrt{1-C^2}}{2}
-\frac{1-\sqrt{1-C^2}}{2}\log_2 \frac{1-\sqrt{1-C^2}}{2}$, where
\begin{eqnarray}\label{Conc}
C=\max(0, 2 \lambda_{max} - \sum_{i=1}^4 \lambda_i), \;\;
\lambda_{max} = \max(\lambda_1,\lambda_2,\lambda_3,\lambda_4).
\end{eqnarray}
Here $\lambda_i$ are the eigenvalues of the following matrix
\begin{eqnarray}
\tilde\rho^{(nm)} = \sqrt{\rho^{nm} (\sigma_y \otimes \sigma_y)(\rho^{nm})^* 
(\sigma_y \otimes \sigma_y)}, \;\;\sigma_y=\left(
\begin{array}{cc}
0&-i\cr
i&0\end{array}\right).
\end{eqnarray}
Hereafter we { consider} the concurrence as a measure of entanglement.

\section{Relay entanglement in symmetrical model} 
\label{Section:relay}

{\subsection{Preliminary remarks and symmetrical initial state }}
{
 In general, the model introduced in Sec.\ref{Section:model} possesses 6 parameters of the initial state (\ref{inSR}-\ref{UR}). However,  
according to the results of Ref. \cite{DZ_2017}, the effect of $\alpha_2$ and $\beta_2$ on entanglement 
is negligible. In addition, we consider the symmetrical model for simplicity.  Thus we set
\begin{eqnarray}\label{sym}
\lambda^R=\lambda^S=\lambda,\;\;\alpha_1=\beta_1=\alpha, \;\;\alpha_2=\beta_2=0.
\end{eqnarray}
In this setting, 
averaging {the} sum $s$ in  (\ref{s}) over the parameters of the sender's and receiver's initial state    reduces  to
\begin{eqnarray}\label{avs}
\big\langle s \big\rangle = {\pi}\int_0^1 d\alpha\int_\frac{1}{2}^1 d \lambda\sin(\alpha \pi) s(\alpha,\lambda) .
\end{eqnarray}
 In the case $N=10$, the numerical analysis shows that  $\big\langle s \big\rangle$  reaches its maximal value at $t=12.238$ (we use dimensionless time setting $D=1$).}

\begin{figure*}
\epsfig{file=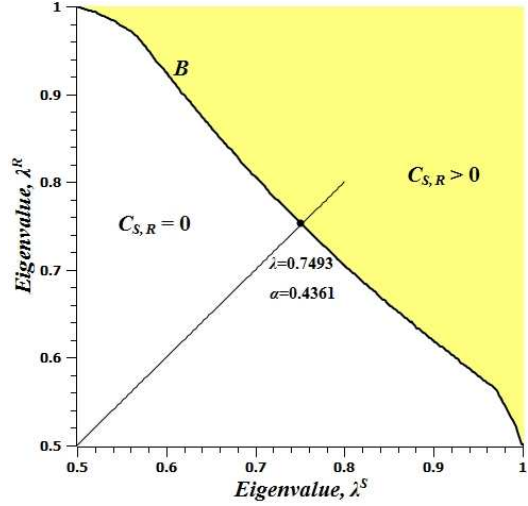,
  scale=0.4
,angle=0
}  
\caption{ The region of zero $SR$-entanglement $C_{S,R}$ for the chain of $N=10$  nodes at the optimal
instant $t=12.238$. The  bisectrix $\lambda^R=\lambda^S$ crosses the boundary $B$ at the point $\lambda=0.7493$ with $\alpha=0.4361$.}
  \label{Fig:zeroE} 
\end{figure*}

As  mentioned in  Introduction, the information transfer between the sender $S$ and receiver $R$ is not related to the   $SR$-entanglement \cite{DZ_2017}. There is a large  region   in the plane of initial eigenvalues $(\lambda^R,\lambda^S)$ such that the 
information transfer occurs without the entanglement  between $R$ and $S$ (for any values of other initial-state parameters)   during the whole 
evolution period from zero to the certain prescribed optimal instant for state registration.  {
For $N=10$, this region of vanishing  $SR$-entanglement, bounded by the curve $B$,  is shown in Fig.\ref{Fig:zeroE}. }

The bisectrix $\lambda^R=\lambda^S$ in  Fig.\ref{Fig:zeroE} crosses the boundary curve $B$ at the point $\lambda=0.7493$ with $\alpha=0.4361$.  { These values have been found in a straightforward way, see Ref.\cite{DZ_2017}. Namely, increasing $\lambda$ along the  bisectrix we find the critical value ($\lambda=0.7493$ ) such that there is no entanglement at $\lambda$ below this value (independently on the value of  $\alpha$) and, at the critical value $\lambda=0.7493$, the  $SR$-entanglement is nonzero only at $\alpha=0.4361$. Thus, the point ($\lambda,\alpha$) =(0.7493,0.4361) on the  bisectrix in Fig.\ref{Fig:zeroE}  is the point of entanglement arising.} 

{\subsection{Partial sums and relay entanglement}
}
Our basic assumption is that the pairwise correlations are  responsible for the state propagation in a quantum system so that the rest of this section is devoted to studying such   combinations of pairwise entanglements which remain valuable during the whole  evolution period. In this case, we can propose  that the entanglement passes from a given entangled pair to another one 
establishing the relay propagation of  entanglement along the spin chain.  

\begin{figure*}\epsfig{file=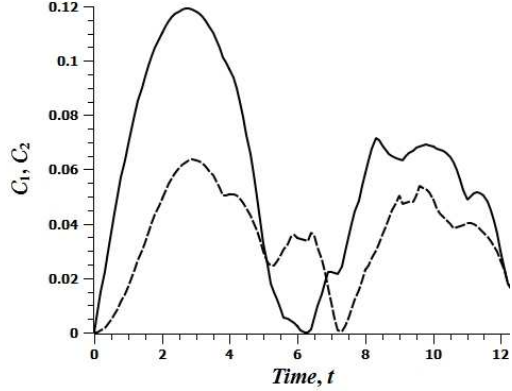,
  scale=0.4
,angle=0
}  
\caption{ The mean concurrences  ${\cal{C}}_1$ (bold line) and ${\cal{C}}_2$ (dashed line) as functions of $t$ 
 at $\lambda=0.7 $ and $\alpha=0$. Here and  below $N=10$.}
  \label{Fig:sum} 
\end{figure*}

Let $C_{i,j}$ be the pairwise concurrence between the $i$th and  $j$th particles. We distribute all $C_{i,j}$ in the groups of concurrences between the equidistant nodes ($|i-j|=m=const$), and consider the mean concurrences
${\cal{C}}_m$  in such groups:
\begin{eqnarray}\label{E1}
&&
{\cal{C}}_m = \frac{1}{N-m}\sum_{i=1}^{N-m} C_{i,i+m},\;\;m=1,2,\dots,N-1,
\end{eqnarray}
as  functions of the initial state parameters ($\lambda$ and $\alpha$) and time $t$. To illustrate the time-behavior of these functions, we depicture  two of them, 
 ${\cal{C}}_1$ and ${\cal{C}}_2$, for the particular values $\lambda=0.7$  and $\alpha=0$  in Fig.\ref{Fig:sum}.
We see that, in general, ${\cal{C}}_1>{\cal{C}}_2 $, but there  is a time interval where the contribution from 
${\cal{C}}_2 $ prevails over the contribution from ${\cal{C}}_1$.
Thus, the entanglement between the nearest neighbors does not always dominate.
This prompts us to introduce the  sum of all pairwise entanglements in the chain 
\begin{eqnarray}\label{SmN}
S = \sum_{i=1}^N{\cal{C}}_i  
\end{eqnarray}
as a measure of overall overlaps of pairwise concurrences. We refer to $S$ as the relay entanglement.
However, {the} relay entanglement does not explore the contribution from different  mean concurrences  ${\cal{C}}_i$.  
To reveal this  effect,    we 
 introduce the partial sums $S_m$
\begin{eqnarray}\label{Sm}
S_m = \sum_{i=1}^m{\cal{C}}_i,\;\;m=1,2,\dots,N-1.
\end{eqnarray}
The relay entanglement $S$ and the partial sums $S_m$ are 
functions of $t$, $\lambda$ and $\alpha$.
%
There is an evident chain of inequalities:
\begin{eqnarray}
S_1\le S_2 \le \dots \le S.
\end{eqnarray}
To characterize the partial sums (which are  functions of three variables),  we consider the maximal values of these sums over 
the time interval $0\le t \le 12.238$ and their minimal values over the time interval  $1\le t \le 12.238$:
\begin{eqnarray}\label{max}
S_m^{max}(\lambda,\alpha)= \max_{0\le t \le =12.238} S_m(\lambda,\alpha,t),\\\label{min}
S_m^{min}(\lambda,\alpha)= \min_{1\le t \le =12.238} S_m(\lambda,\alpha,t).
\end{eqnarray}
In the later case the 
shift by 1 in the time interval is conventional, we use it to remove the initial evolution interval because all pairwise entanglements  are zero at $t=0$ due to  our choice of the initial state.  The contribution  from the $m$th group is negligible if  $\displaystyle\frac{|S_{m-1}^{max}-S_m^{max}|}{S_{m-1}^{max}} \ll 1 $ and  
$\displaystyle\frac{|S_{m-1}^{min}-S_m^{min}|}{S_{m-1}^{min}} \ll 1 $. 

To depicture the behavior of two-parametric  functions $S_m^{max}$ and $S_m^{min}$, we introduce the   mean value and  the root-mean-square deviation with respect to the parameters $\alpha$ and $\lambda$. Thus, for any 
function $S$ of the two parameters $\lambda$ and $\alpha$, the mean value and root-mean-square deviation with respect to $\alpha$ read:
\begin{eqnarray}\label{mean}
&&
\big \langle S \big\rangle_\alpha(\lambda) =\frac{\pi}{2} \int\limits_0^1 d\alpha \sin(\alpha\pi) S(\lambda,\alpha)
,\\\label{dev}
&&
\delta_\alpha S(\lambda) =\sqrt{\Big\langle (S(\lambda,\alpha)- \big\langle S \big\rangle_\alpha(\lambda))^2\Big\rangle_\alpha}.
\end{eqnarray}
Similarly, the mean value and root-mean-square deviation with respect to $\lambda$ read:
\begin{eqnarray}\label{mean2}
&&
\big \langle S \big\rangle_\lambda(\alpha) = 2\int\limits_{1/2}^1 d\lambda S(\lambda,\alpha)
,\\\label{dev2}
&&
\delta_\lambda S(\alpha) =\sqrt{\Big\langle (S(\lambda,\alpha)- \big\langle S \big\rangle_\lambda(\alpha))^2\Big\rangle_\lambda}.
\end{eqnarray}
Now we can estimate:
\begin{eqnarray}\label{alpha0}
 &&
 \big \langle S \big\rangle_\alpha - \delta_\alpha S \lesssim S\lesssim \big \langle S \big\rangle_\alpha +
 \delta_\alpha S,\\\label{lam0}
 &&
 \big \langle S \big\rangle_\lambda - \delta_\lambda S \lesssim S\lesssim\big \langle S \big\rangle_\lambda + \delta_\lambda S .
\end{eqnarray}
Therefore we can represent  $S$ graphically 
as the mean values $\big \langle S \big\rangle_\alpha$ (or $\big \langle S \big\rangle_\lambda$)
supplemented by the root-mean-square deviations $\delta_\alpha S$ (or $\delta_\lambda S$)  as the error-bars. 

\subsection{Analysis of partial sums}
\label{Section:analysis}
\begin{figure*}
\subfloat[]{\includegraphics[scale=0.4,angle=0]{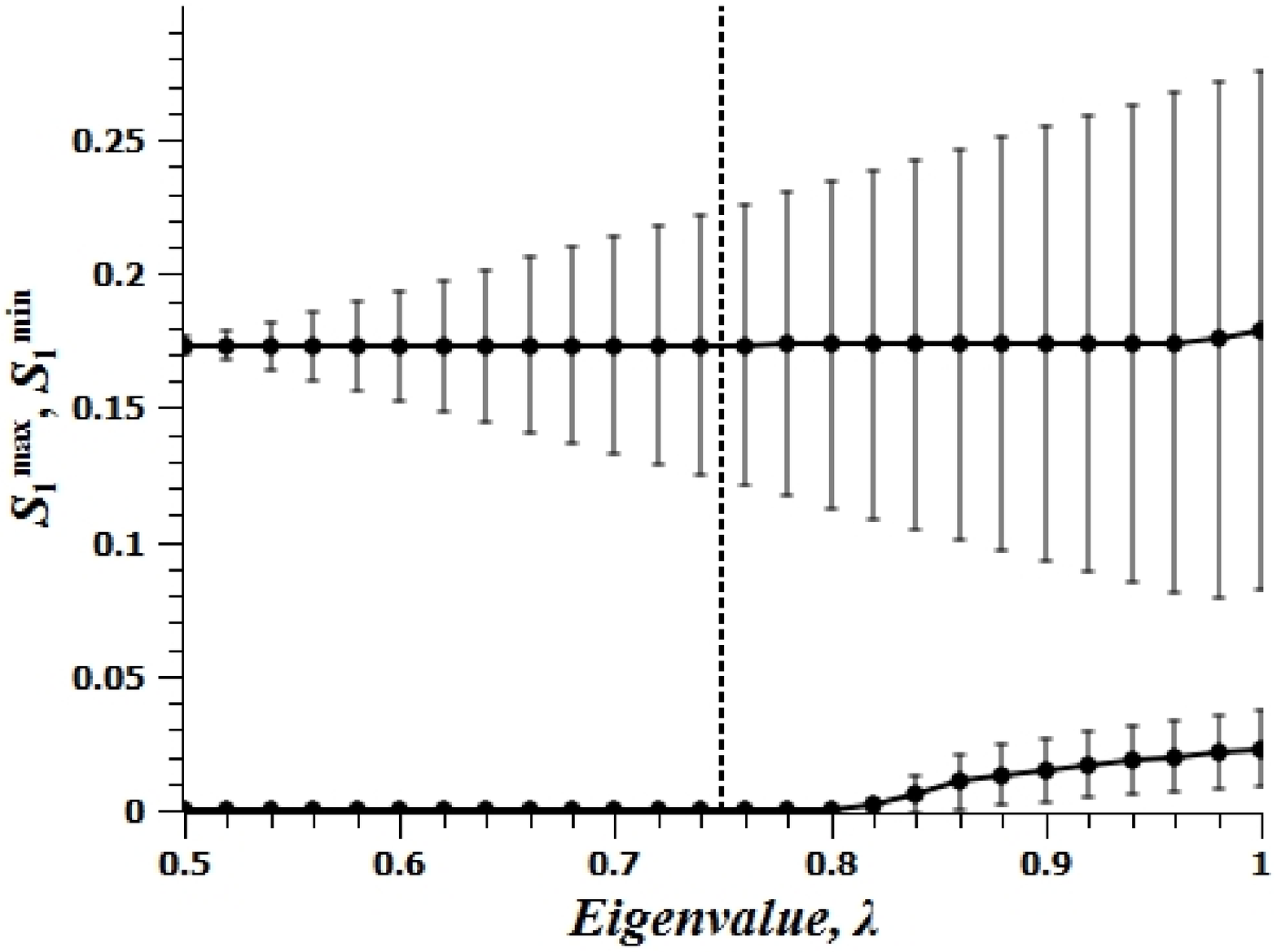
}}
\subfloat[]{\includegraphics[scale=0.4,angle=0]{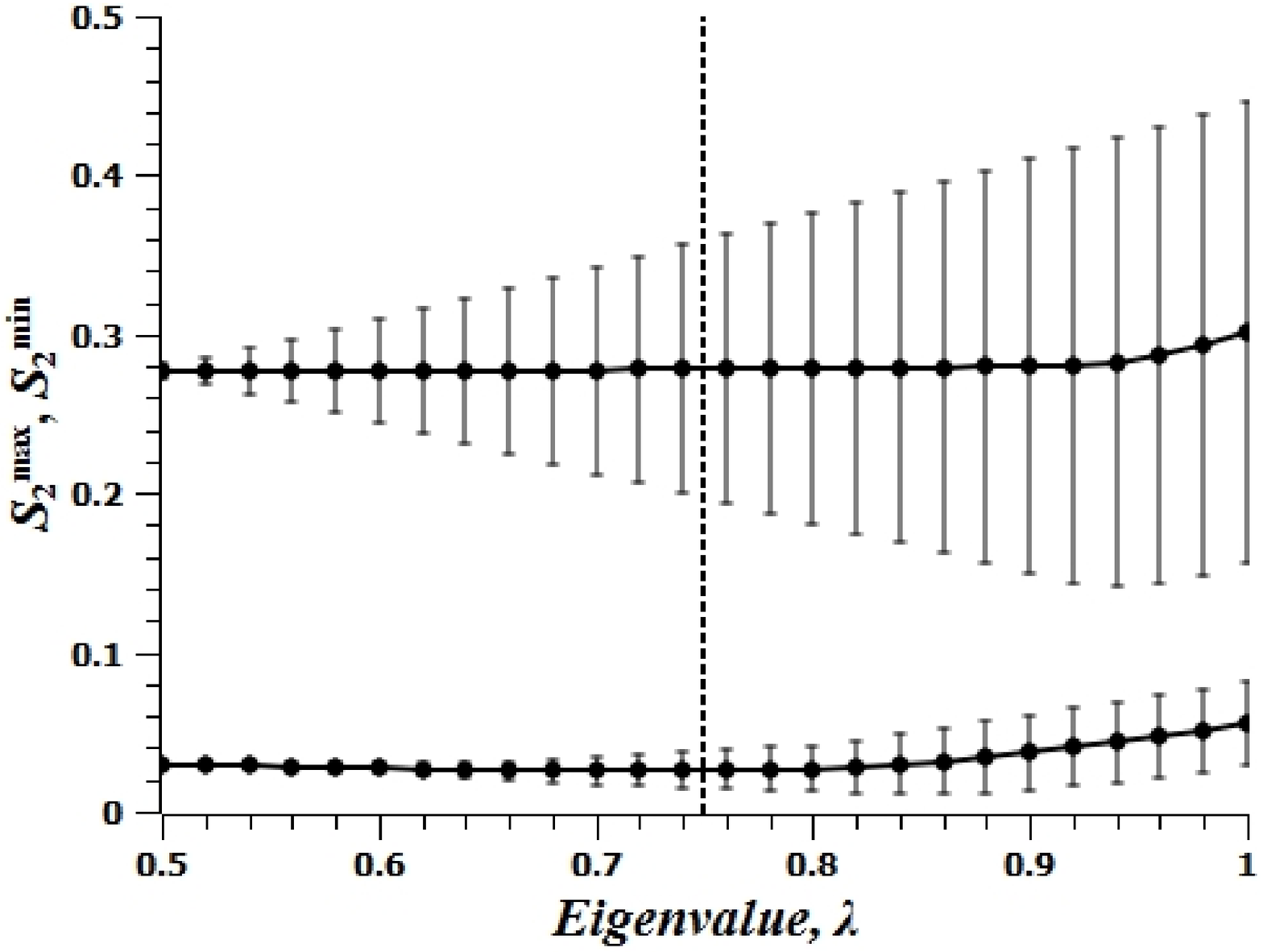
}}\\
\subfloat[]{\includegraphics[scale=0.4,angle=0]{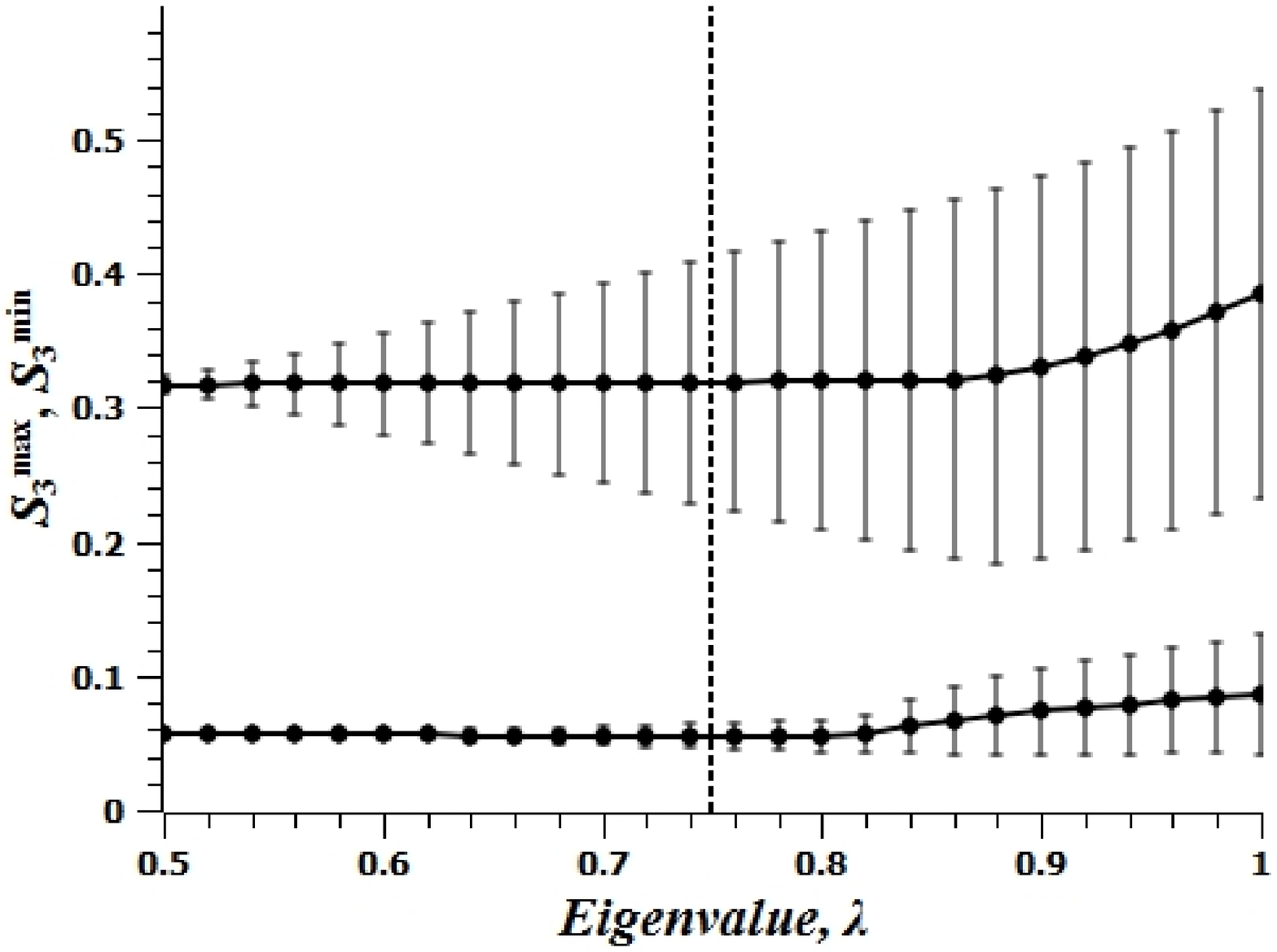
}}
\subfloat[]{\includegraphics[scale=0.4,angle=0]{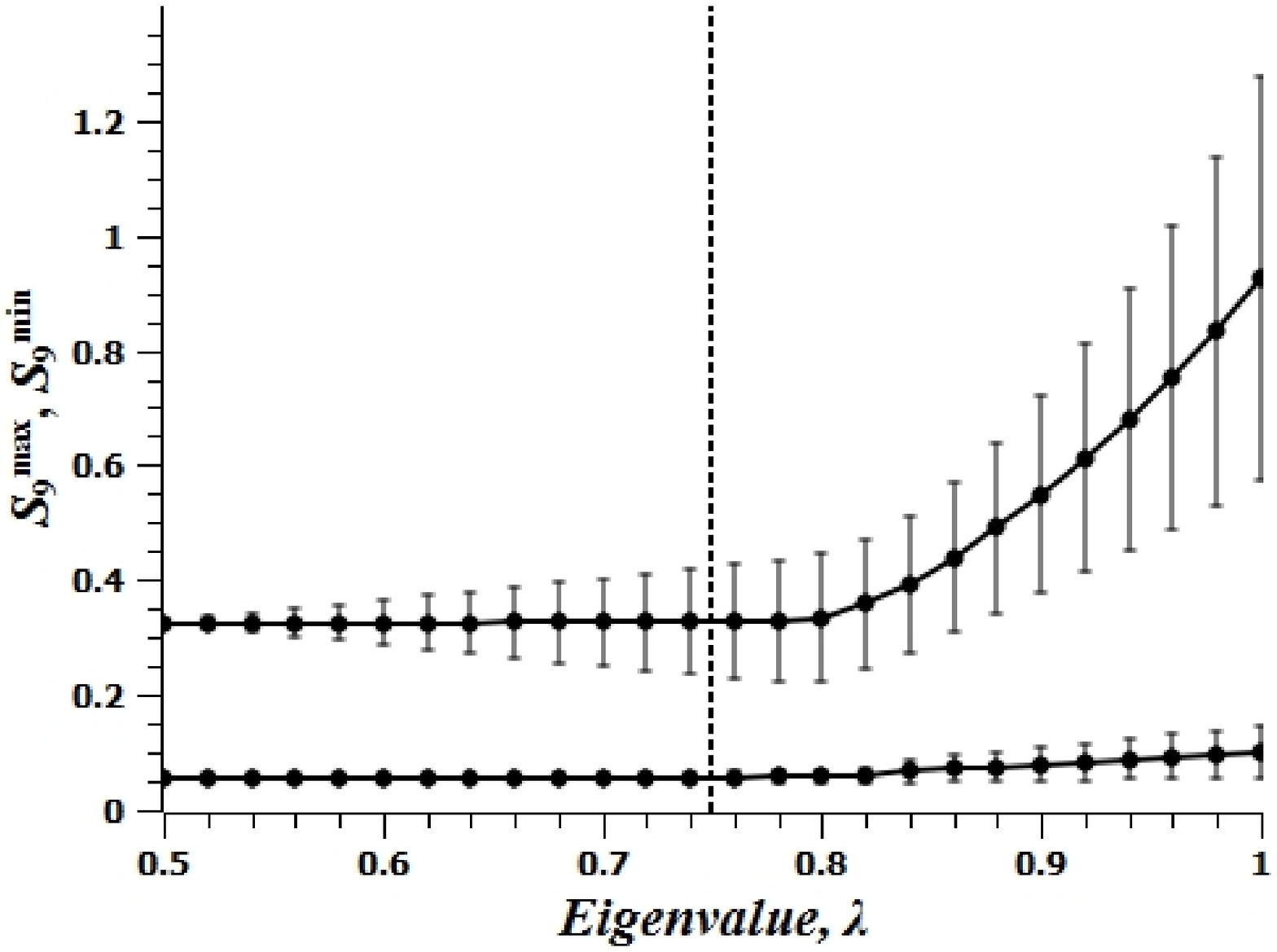
}}
\caption{The maximum $S_m^{max}\approx \big \langle S_m^{max} \big\rangle_\alpha \pm \delta_\alpha S^{max}$ (upper curve) and minimum
$S_m^{min}\approx\big \langle S_m^{min} \big\rangle_\alpha \pm \delta_\alpha S^{min}$ (lower curve) as functions of $\lambda$.
a) $m=1$, b) $m=2$, c) $m=3$, d) $m=9$. The vertical dotted  lines correspond to the intersection point of the bisectrix and boundary curve in Fig.\ref{Fig:zeroE} ($\lambda=0.7493$). }
  \label{Fig:avrAllalpha} 
\end{figure*}

\begin{figure*}
\subfloat[]{\includegraphics[scale=0.4,angle=0]{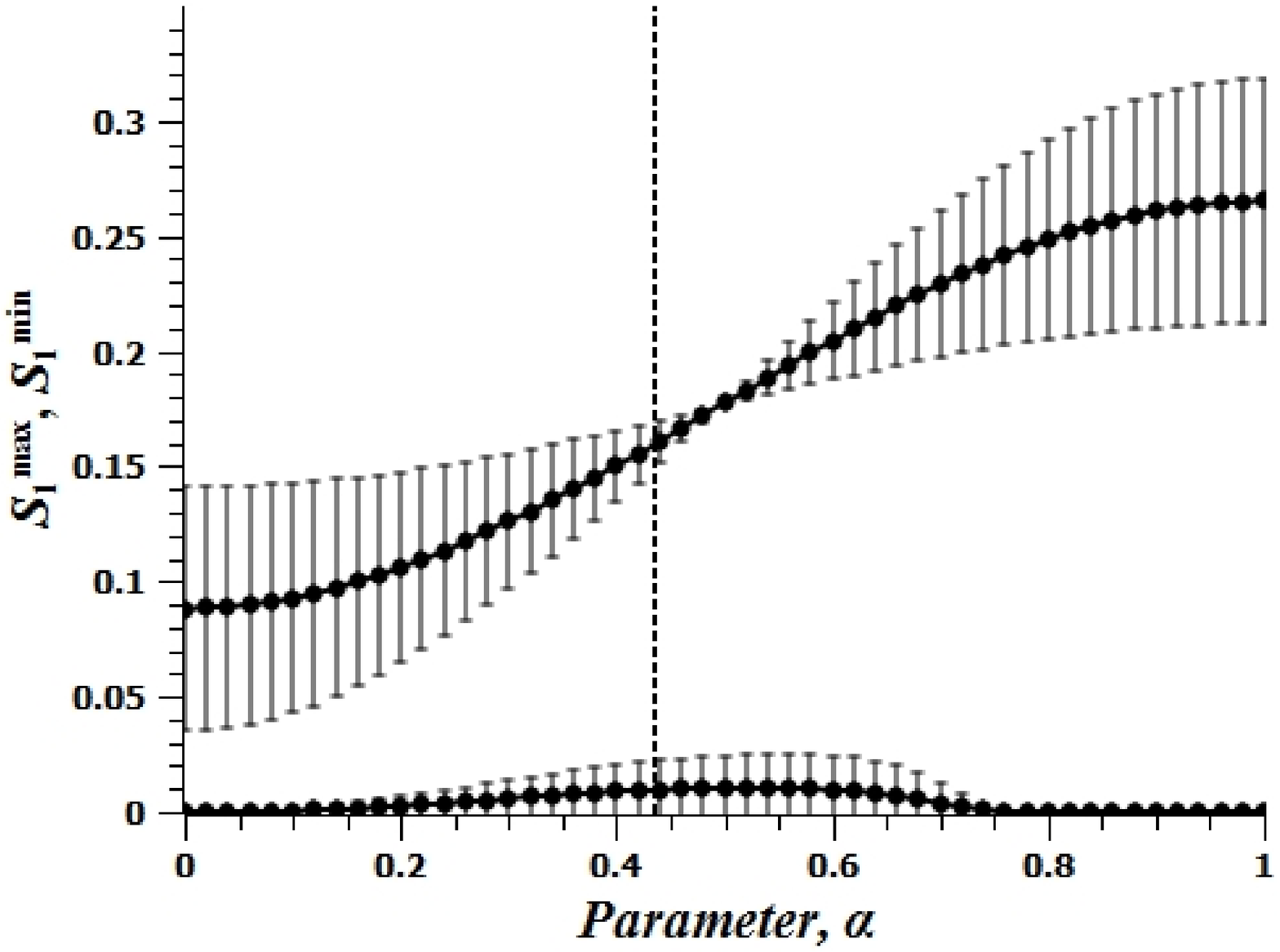
}}
\subfloat[]{\includegraphics[scale=0.4,angle=0]{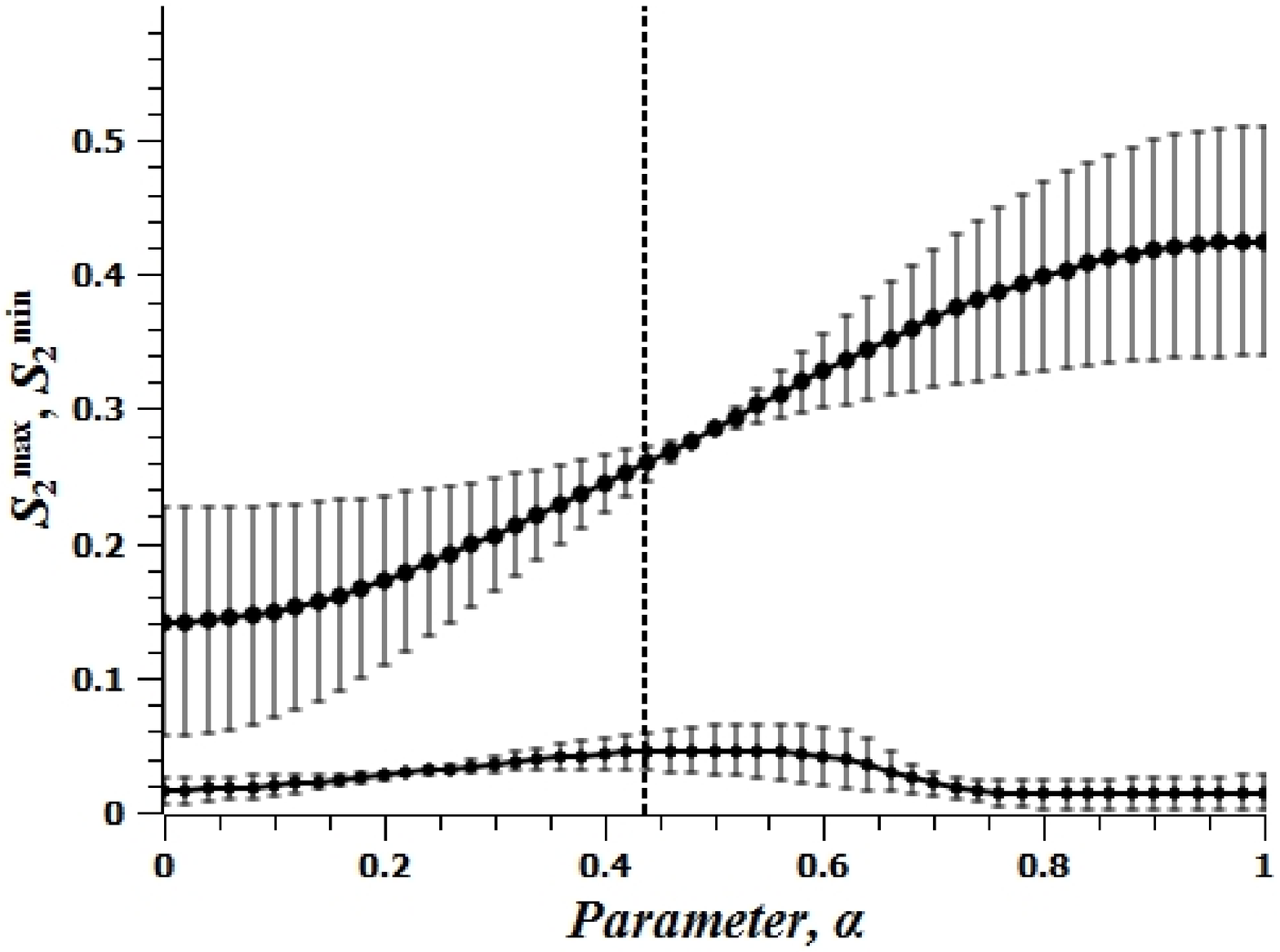
}}\\
\subfloat[]{\includegraphics[scale=0.4,angle=0]{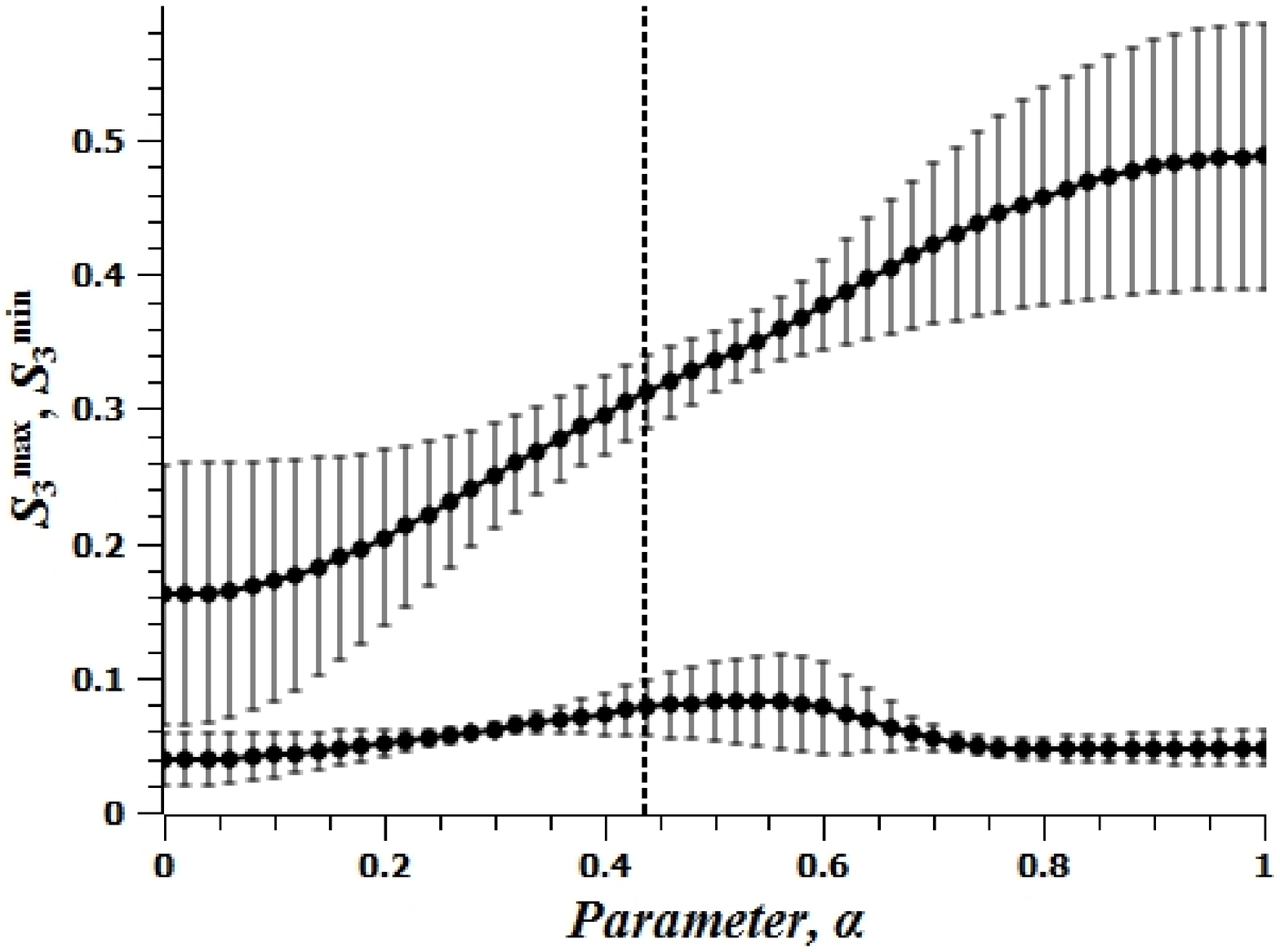
}}
\subfloat[]{\includegraphics[scale=0.4,angle=0]{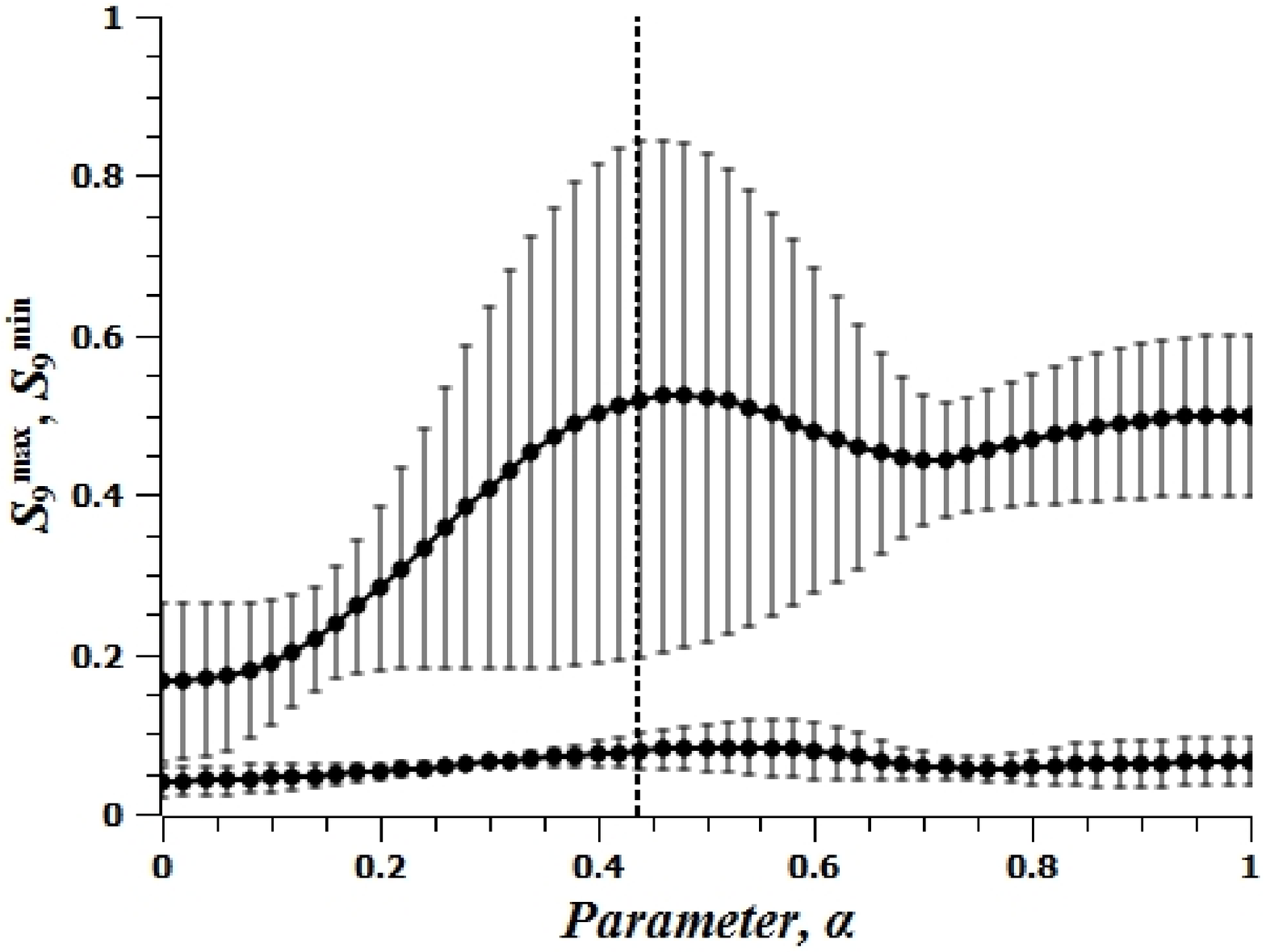
}}
\caption{The maximum  $S_m^{max}\approx  \big \langle S_m^{max} \big\rangle_\lambda\pm \delta_\lambda S^{max}$ (upper curve) and the minimum
$S_m^{min}\approx  \big \langle S_m^{min} \big\rangle_\lambda\pm \delta_\lambda S^{min}$ (lower curve) as functions of $\alpha$. 
a) $m=1$, b) $m=2$, c) $m=3$, d) $m=9$. The vertical dotted lines correspond to the intersection point of the bisectrix and boundary curve in Fig.\ref{Fig:zeroE} 
($\alpha=0.4361$)}
  \label{Fig:avrAlllambda} 
\end{figure*}

Accordingly, the  functions $ S_m^{max} $ and  $ S_m^{min} $ ($m=1,2,3,9$) in terms of the mean values and root-mean-square deviations with respect to $\alpha$ are shown in Fig.\ref{Fig:avrAllalpha}, and those functions in terms of 
the mean values and root-mean-square deviations with respect to $\lambda$ are depictured in Fig.\ref{Fig:avrAlllambda}.
Both these figures show that  the functions $S_m^{max}$
and $S_m^{min}$ are significantly different for   $m=1,2,3$, and therefore,  the groups ${\cal{C}}_{m}$, $m=1,2,3$, are the most important in the correlation propagation.
 For $m>3$, this difference is much less significant, except the intervals   $\lambda\gtrsim 0.8$  (Fig.\ref{Fig:avrAllalpha}, compare graphs c,d) and  $0.2\lesssim\alpha\lesssim 0.7$ (Fig.\ref{Fig:avrAlllambda},  compare graphs c,d). Thus, all the concurrence groups ${\cal{C}}_m$, $m=1,\dots,9$, contribute to the entanglement propagation if the initial state parameters  are inside of these intervals  and only few of them contribute to the entanglement propagation otherwise.

 Fig. \ref{Fig:avrAllalpha}a,b shows that the maxima of   partial
 sums $S_m$ with $m=1,2$ are almost indepdendent on $\lambda$, but their dependence on $\alpha$ increases with an increase in $\lambda$, which is indicated by the length of the error-bars. The later statement is correct for any $m$ as shown in this figure. With an increase in $m$, these maxima become more  $\lambda$-dependent especially  for $\lambda\gtrsim 0,8$.

  Fig. \ref{Fig:avrAlllambda} is significantly different.   Fig. \ref{Fig:avrAlllambda}a-c shows that the maximum of  partial sums $S_m$ increases with $\alpha$ for small $m$. The dependence on $\lambda$ (reflected by the error bars) is minimal in the middle of $\alpha$-interval for small $m$. However, with an increase in $m$, the $\alpha$-dependence becomes more complicated  and $\lambda$-dependence becomes maximal at the    
   middle of the $\alpha$-interval, $0.2\lesssim \alpha \lesssim 0.7$. The largest root-mean-square deviation corresponds to $\alpha\sim 0.4361$ (the crosspoint of the bisectrix and the boundary $B$  in Fig.\ref{Fig:zeroE}).

\section{Clusters of entangled particles}
\label{Section:clusters}
As a problem directly related to the relay entanglement, we consider the formation of   clusters of entangled particles, i.e., such clusters that all the  pairwise entanglements inside of each of them  are significant. 

\begin{figure*}
\epsfig{scale=0.8,angle=0,file=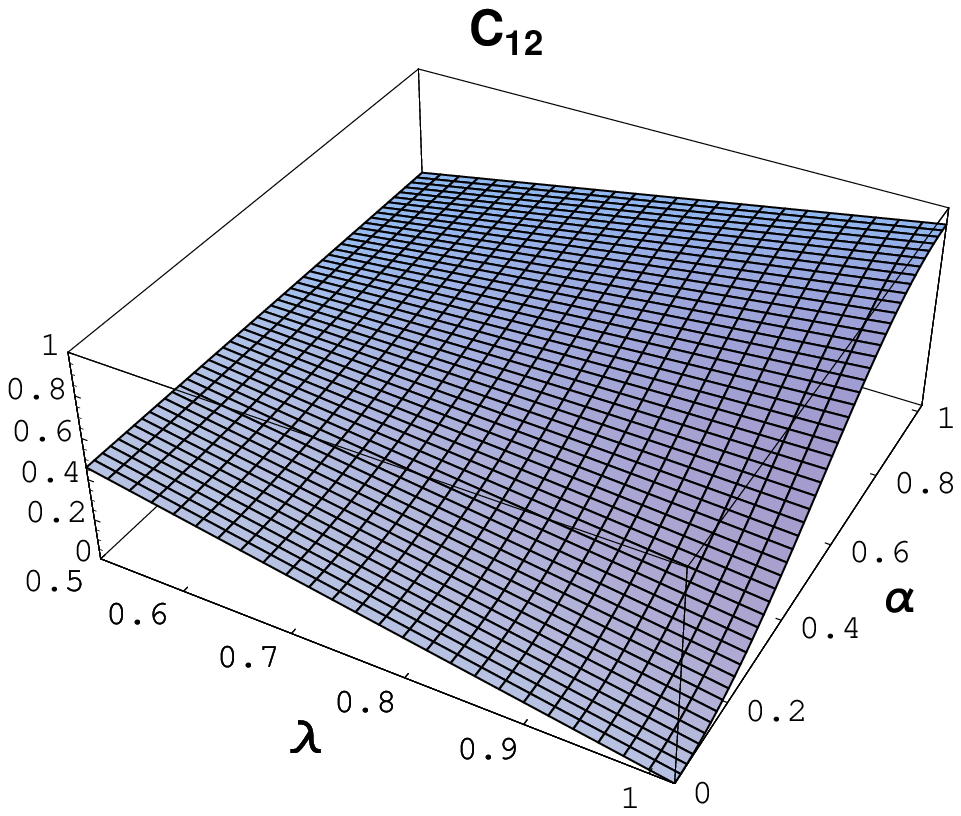
}
\epsfig{scale=0.8,angle=0,file=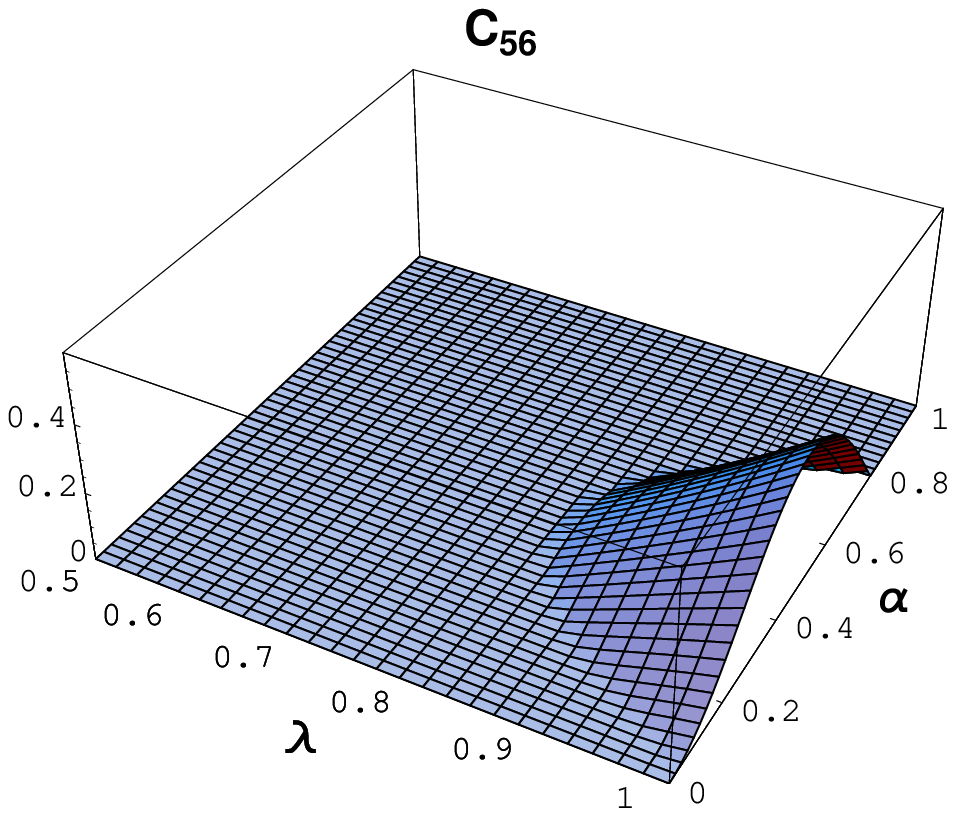
}
\caption{The concurrences $C_{12}$ and $C_{56}$ over the plane  $(\lambda,\alpha)$. }
  \label{Fig:C1256} 
\end{figure*}

Before proceed to study the entanglement in clusters we notice that the symmetry of considered spin chain (the even-number chain, $N=10$) causes the week concurrence in the middle pair, $C_{56}$.
For comparison, we show $C_{12}$ and $C_{56}$ over the plane $(\lambda,\alpha)$ in Fig.\ref{Fig:C1256}.  Unlike  $C_{12}$, the concurrence $C_{56}$ vanishes over the large part of its domain. This affects  the characteristics of those clusters which include the 5th and 6th nodes.

\subsection{Geometric average of pairwise entanglements as measure of quantum correlations in cluster}

To characterize the quantum correlations in a cluster of $M$ particles  {numbered} $i$, $i+1$, $\dots$, $i+M-1$ we introduce the geometric mean of all pairwise concurrences in this cluster:
\begin{eqnarray}\label{E21}
P_{M,i}(\lambda,\alpha,t) =
\left(\displaystyle  \prod_{n-k\le M-1}C_{i+k,i+n} 
\right)^{\frac{1}{\left(M\atop 2\right)}},\;\;\left(M\atop 2\right) = \frac{M!}{2! (M-2)!} = \frac{M(M-1)}{2},
\end{eqnarray}
where $M$ is the number of spins in the cluster (the size of the cluster), and $i$ is the position of the first node of the cluster. Thus, the cluster involves spins from $i$th to $(i+M-1)$th.
The geometric mean  $P_{M,i}$ gets large value  at some  instant if all the concurrences $C_{i,j}$, $i<j\le i+M$, are  large at this  instant.  
This function is localized in time. Generally, its non-zero $t$-support gets narrower with an increase in $M$ and $i$; however, this rule is not strict as is shown below in Fig.\ref{Fig:width}.  To estimate and compare the geometric mean $P_{M,i}$  in  different clusters, we consider the maximum of $P_{M,i}$ over the selected time interval:
\begin{eqnarray}
P_{M,i}^{max}(\lambda,\alpha) = \max_{0\le t \le 12.238}P_{M,i}(\lambda,\alpha,t).
\end{eqnarray}
Similar to Sec.\ref{Section:relay}, we characterize the function of two variables $P_{M,i}^{max}$ using the mean value and root-mean-square deviation with respect to the parameters  $\alpha$ and $\lambda$ defined in formulas (\ref{mean}-\ref{dev2}).

\begin{figure*}
\subfloat[]{\includegraphics[scale=0.3,angle=0]{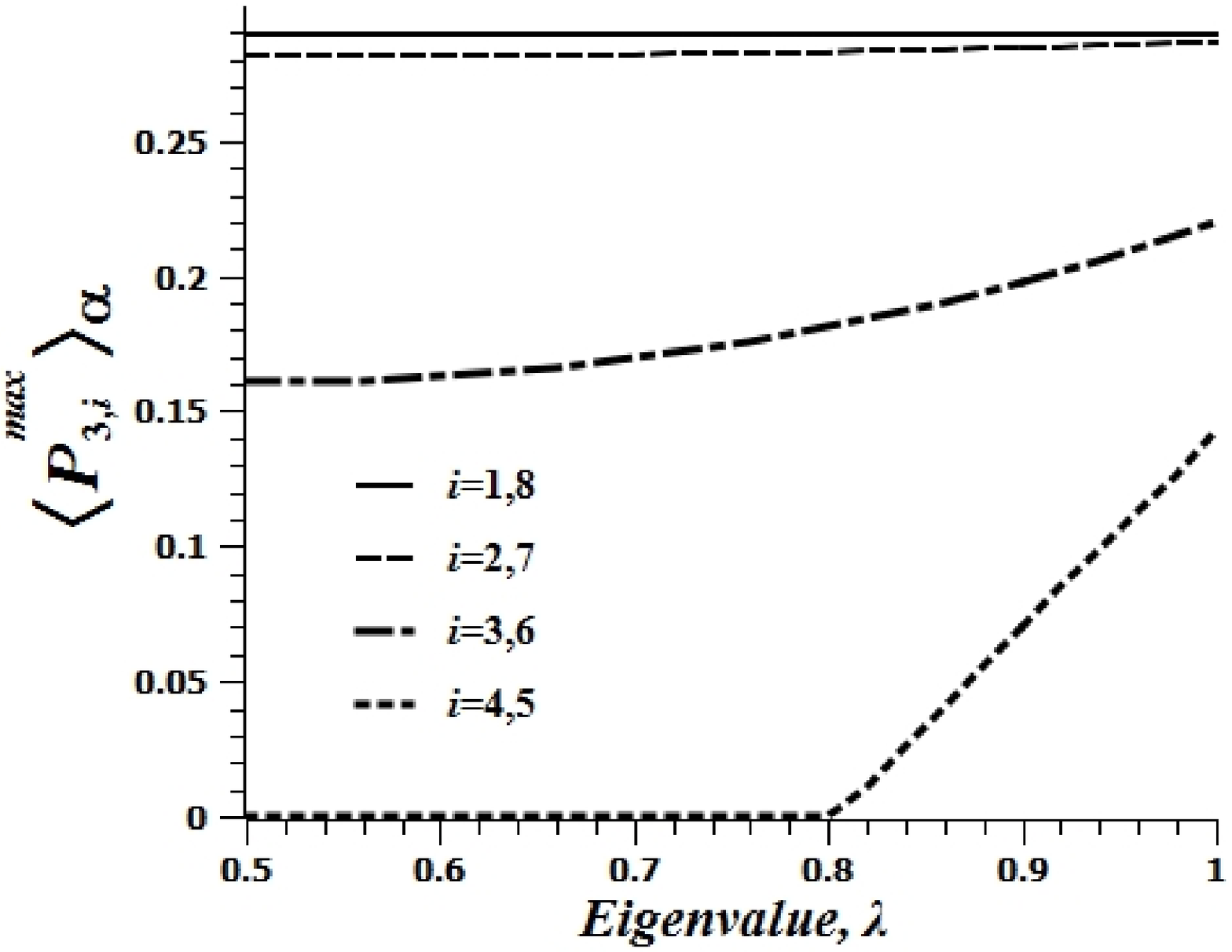
}}
\subfloat[]{\includegraphics[scale=0.3,angle=0]{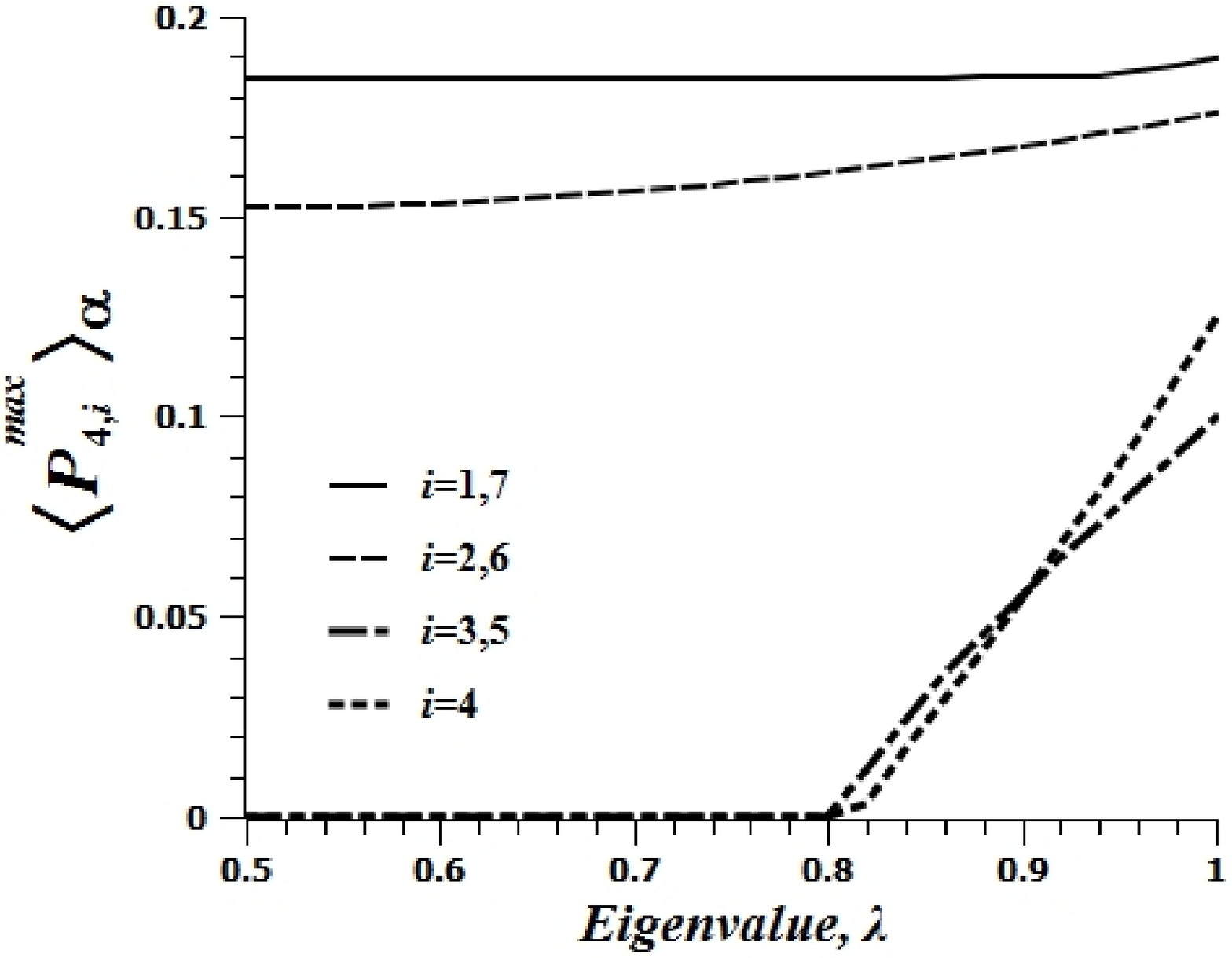
}}
\subfloat[]{\includegraphics[scale=0.3,angle=0]{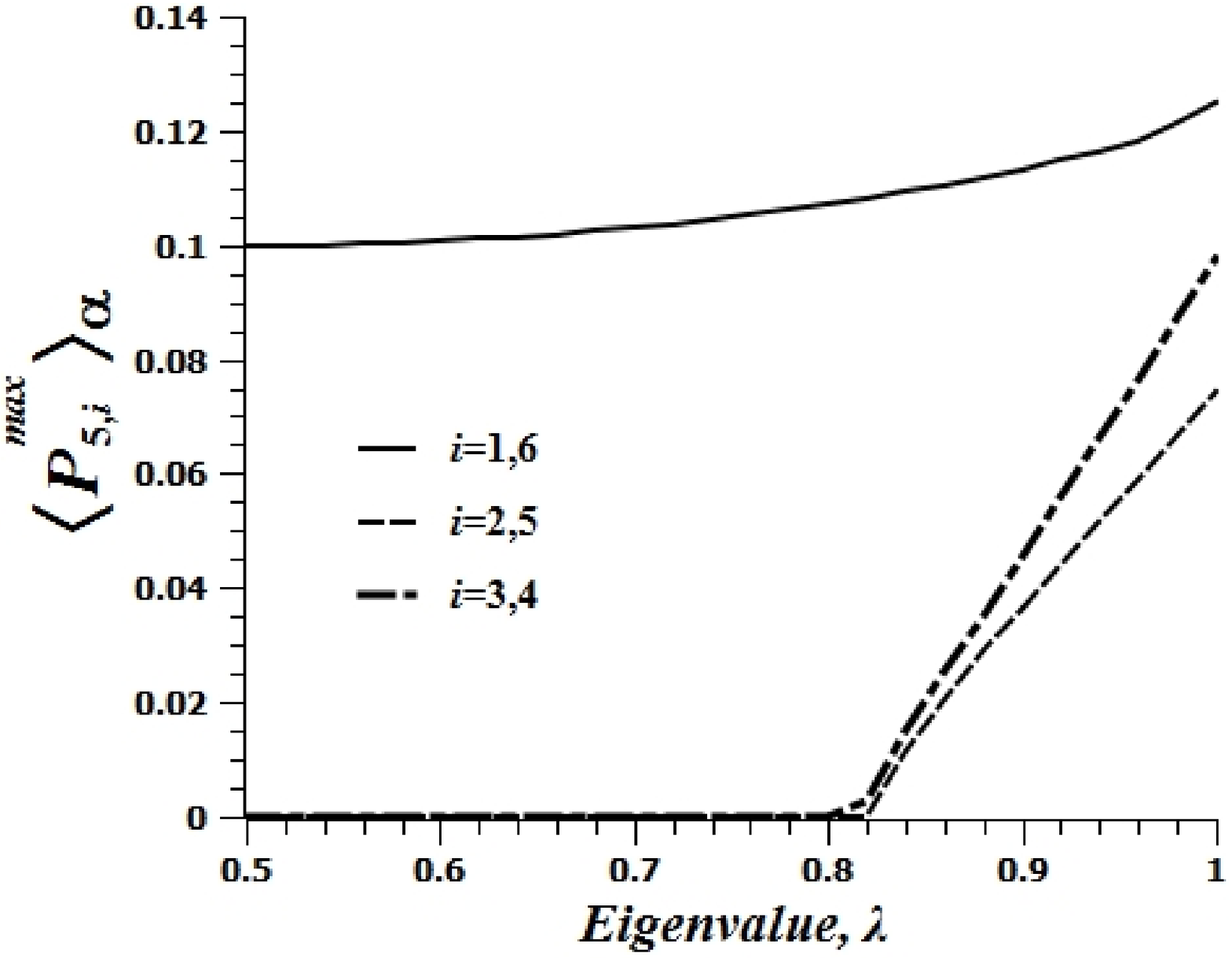
}}\\
\subfloat[]{\includegraphics[scale=0.3,angle=0]{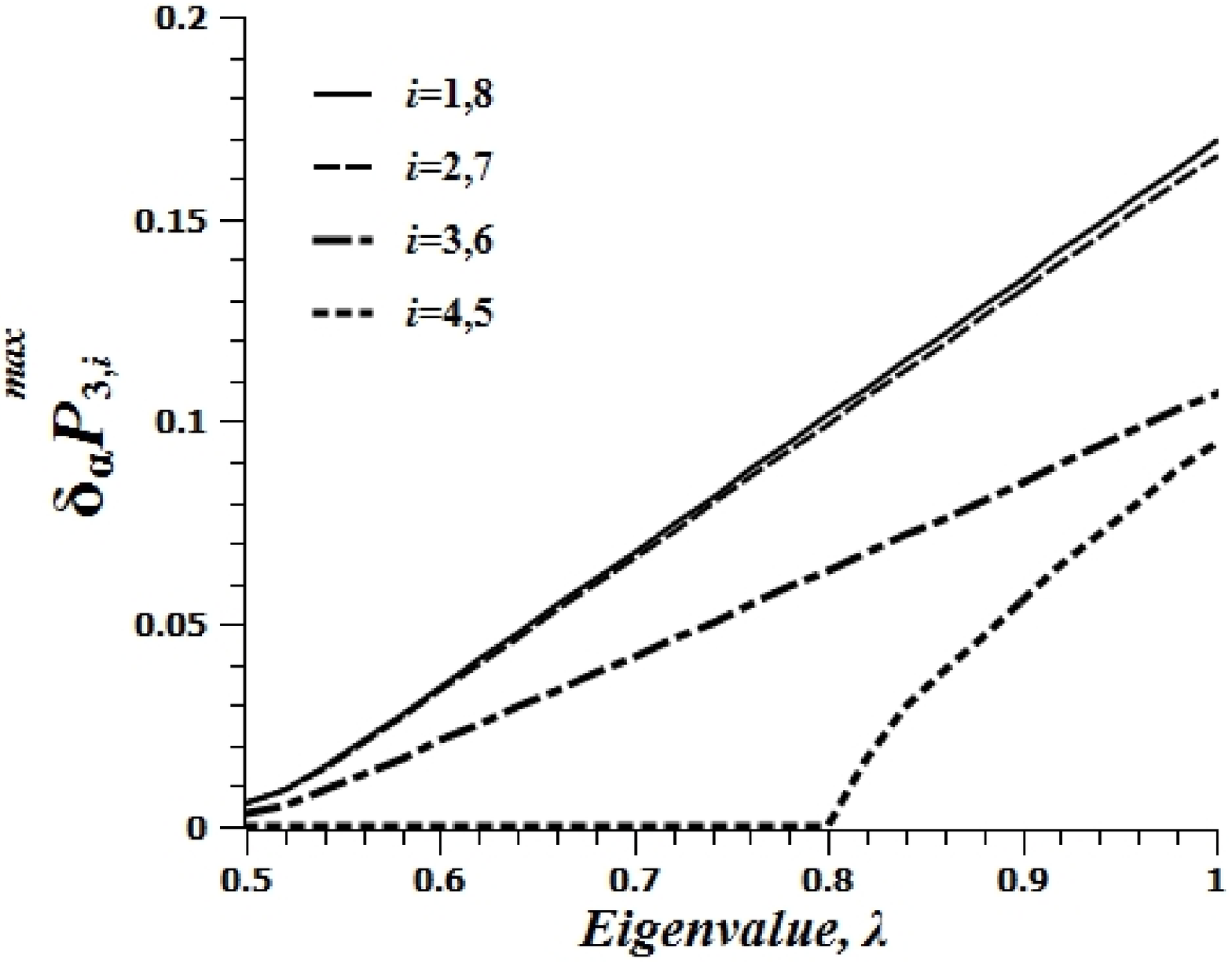
}}
\subfloat[]{\includegraphics[scale=0.3,angle=0]{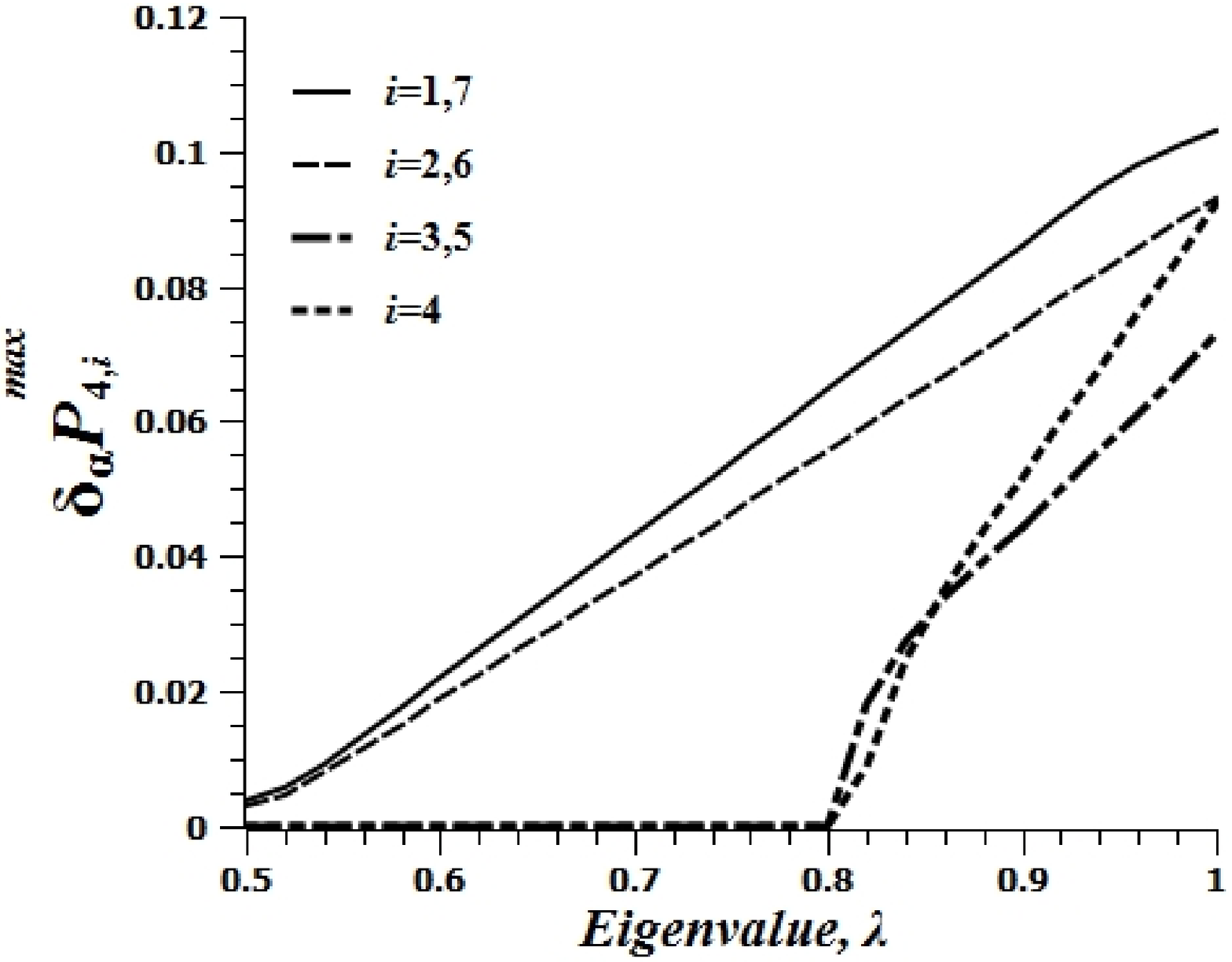
}}
\subfloat[]{\includegraphics[scale=0.3,angle=0]{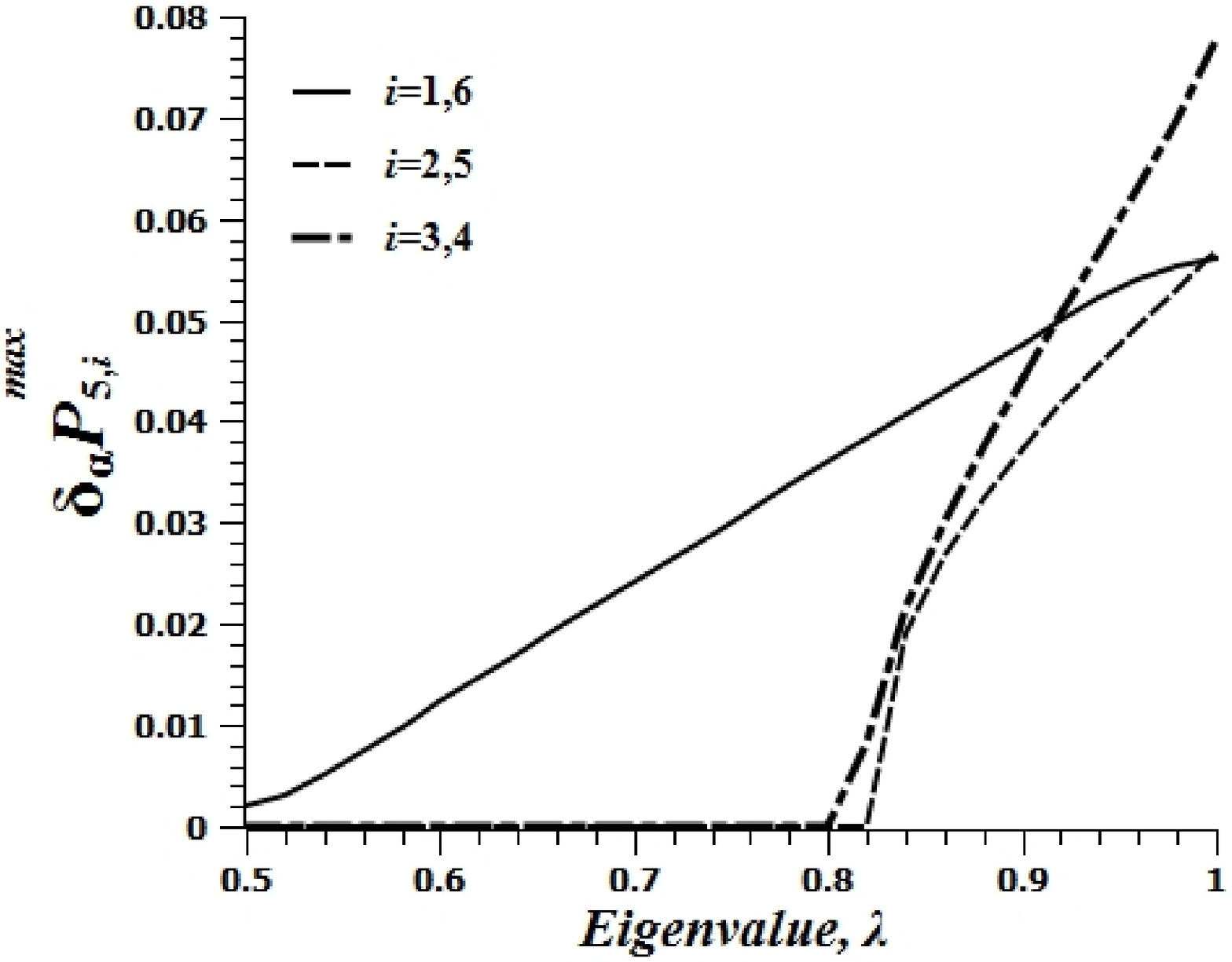
}}
\caption{The mean values  $\big\langle P_{M,i}^{max}\big\rangle_\alpha$ as  functions of $\lambda$:
a) $M=3$, $i=1,\dots,8$, b) $M=4$, $i=1,\dots,7$, c) $M=5$, $i=1,\dots,6$.
The root-mean-square deviations   $\delta_\alpha  P_{M,i}^{max}$ as  functions of $\lambda$:
d) $M=3$, $i=1,\dots,8$, e) $M=4$, $i=1,\dots,7$, f) $M=5$, $i=1,\dots,6$. }
  \label{Fig:avrAll} 
\end{figure*}

\begin{figure*}
\subfloat[]{\includegraphics[scale=0.3,angle=0]{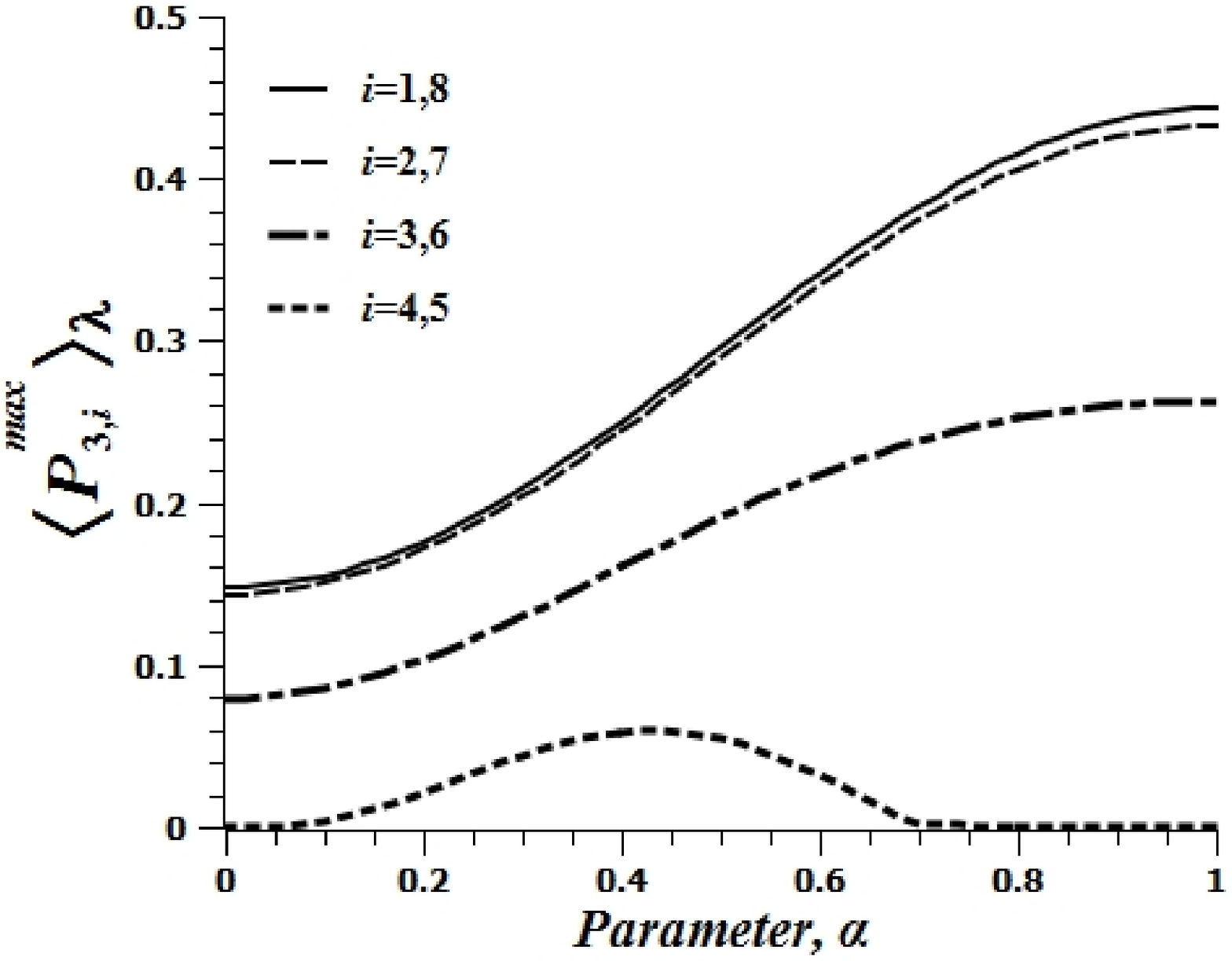
}}
\subfloat[]{\includegraphics[scale=0.3,angle=0]{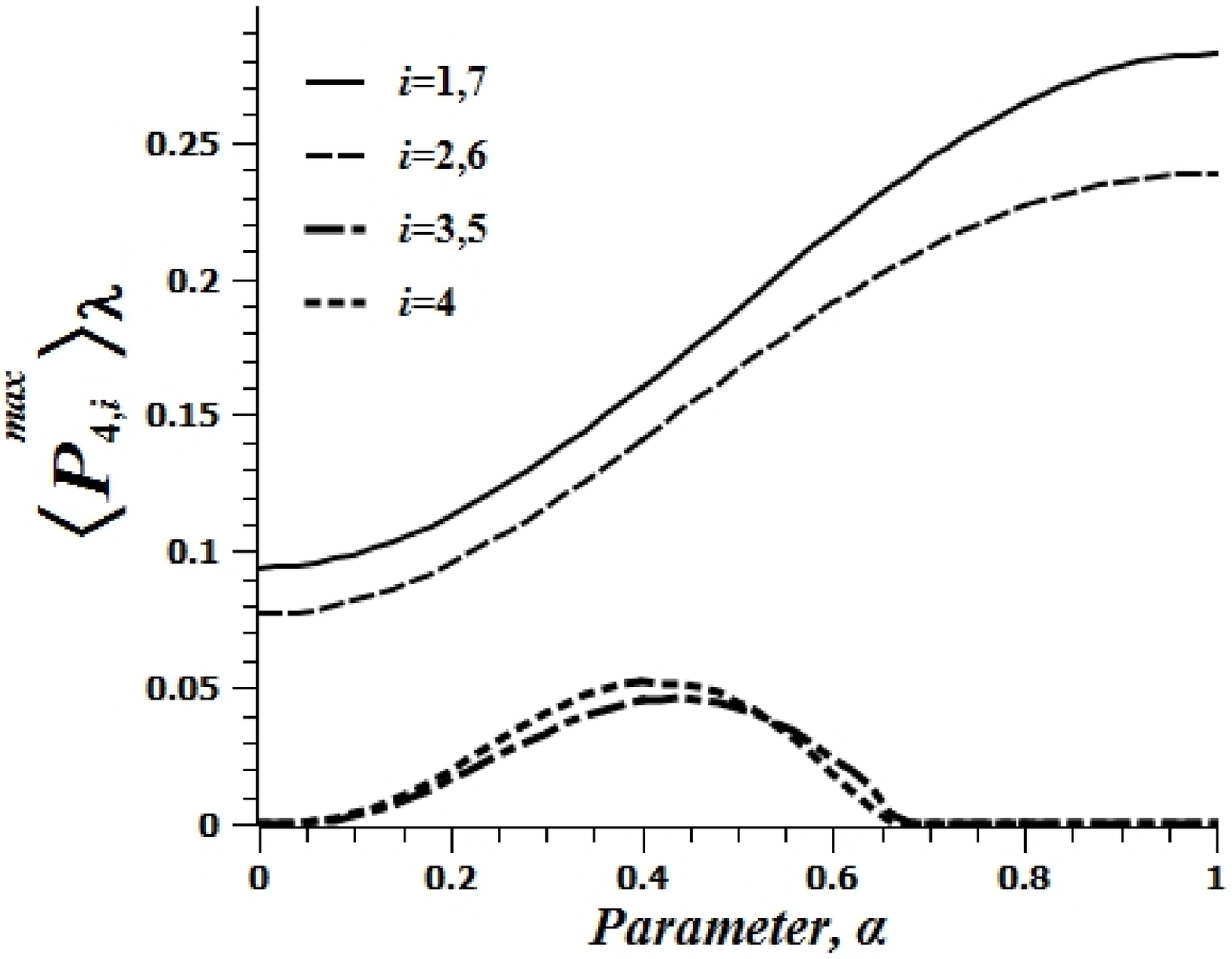
}}
\subfloat[]{\includegraphics[scale=0.3,angle=0]{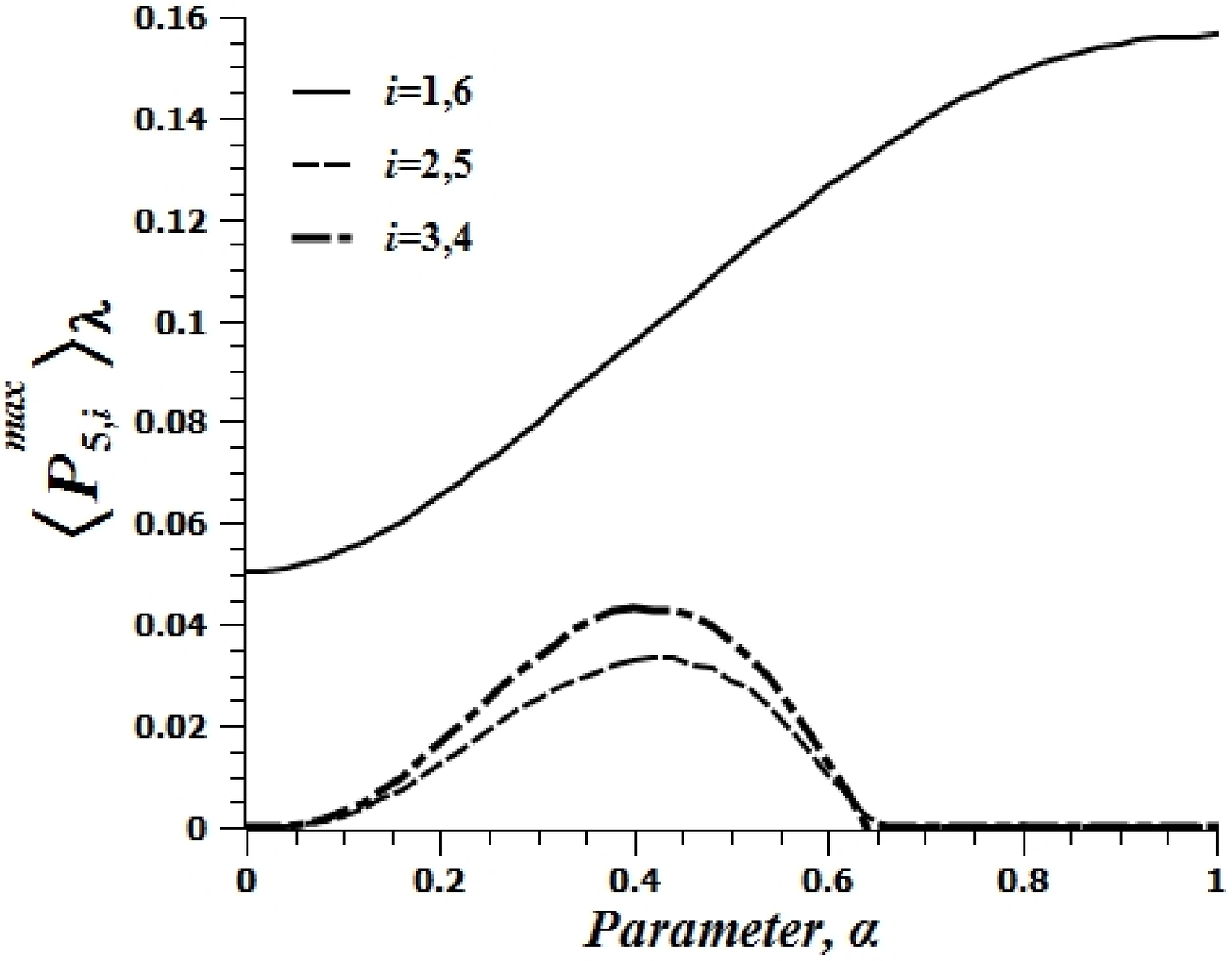
}}\\
\subfloat[]{\includegraphics[scale=0.3,angle=0]{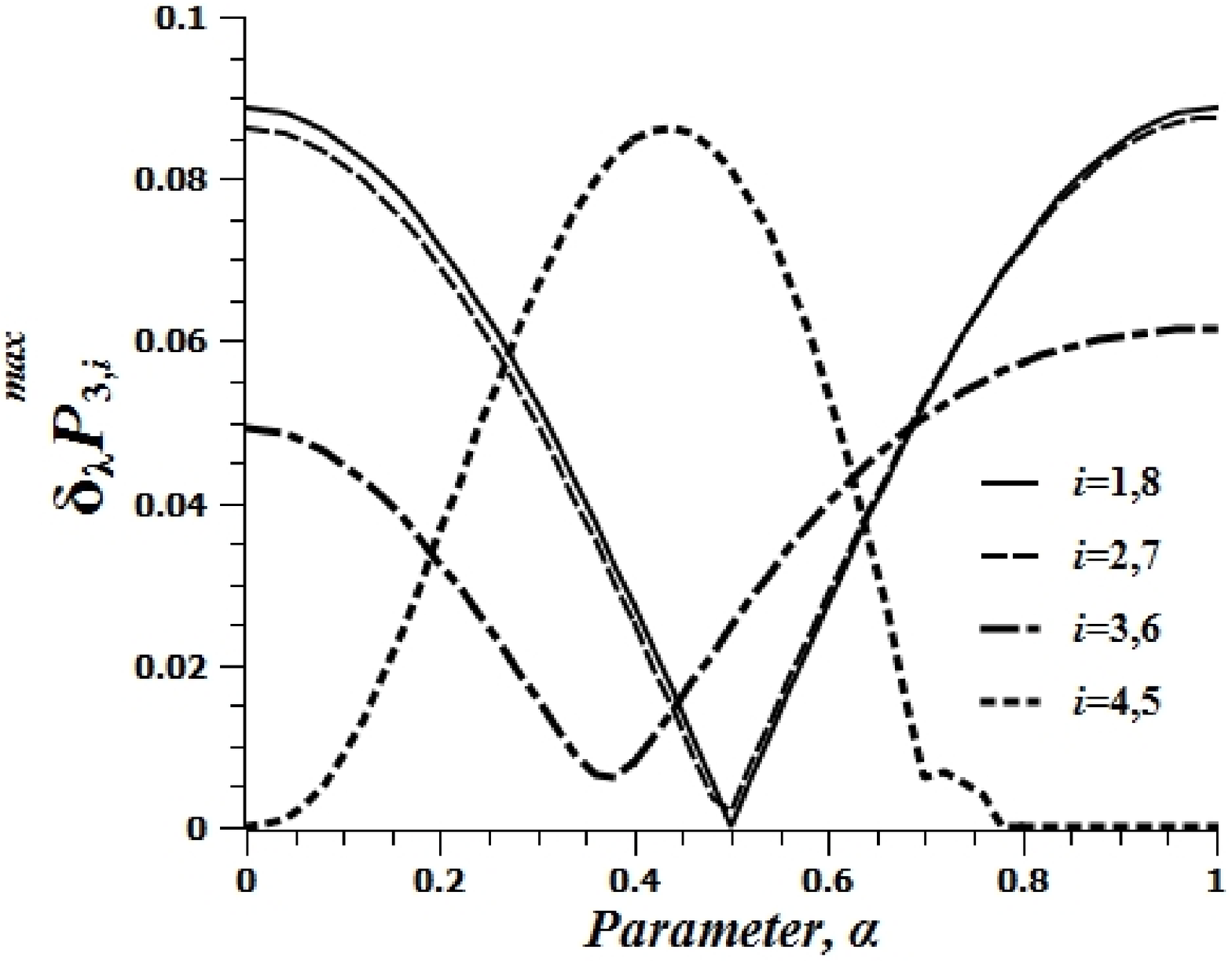
}}
\subfloat[]{\includegraphics[scale=0.3,angle=0]{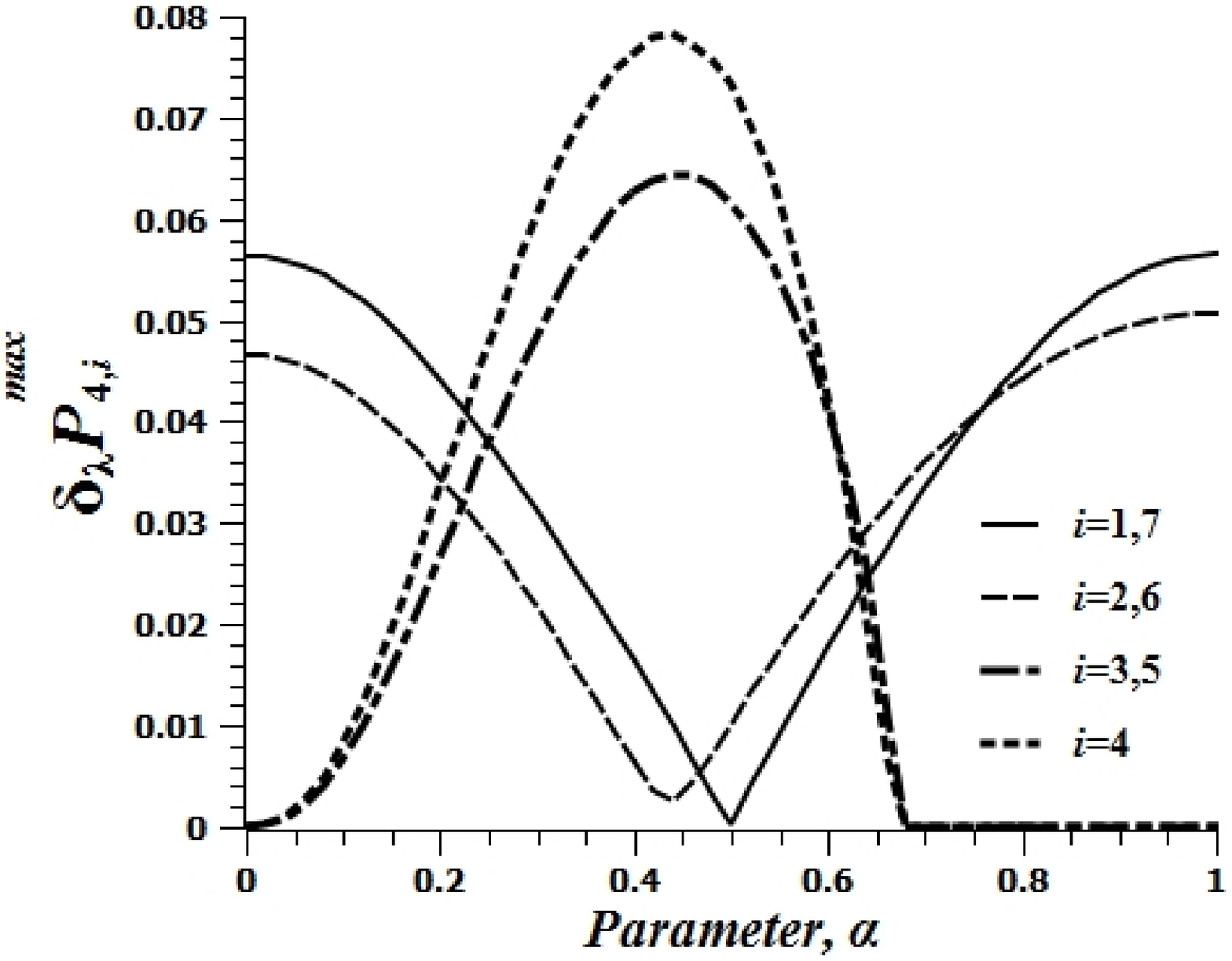
}}
\subfloat[]{\includegraphics[scale=0.3,angle=0]{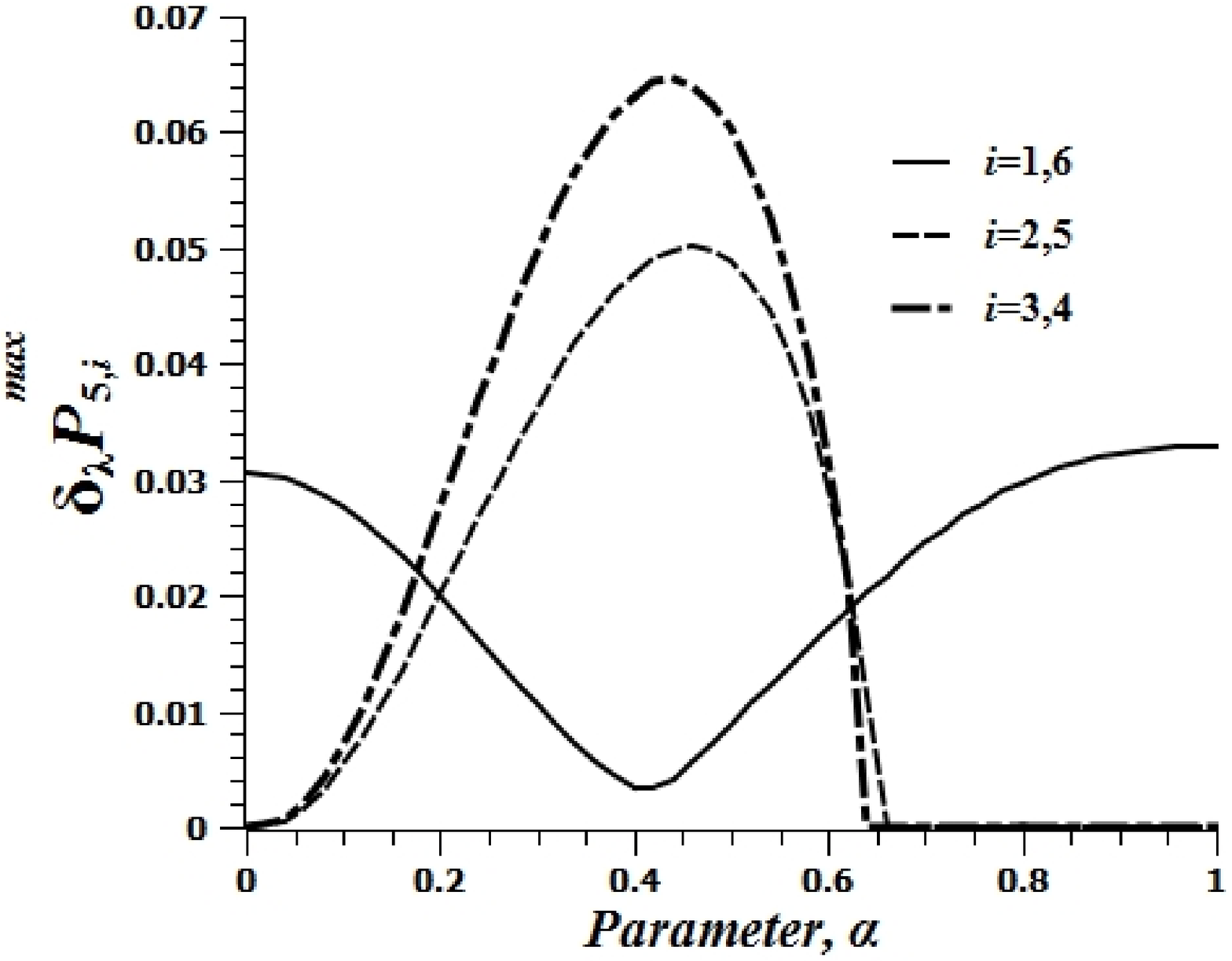
}}
\caption{The mean values  $\big\langle P_{M,i}^{max}\big\rangle_\lambda$ as  functions of $\alpha$.
a) $M=3$, $i=1,\dots,8$, b) $M=4$, $i=1,\dots,7$, c) $M=5$, $i=1,\dots,6$. The root-mean-square deviations   $\delta_\lambda P_{M,i}^{max}$ as  functions of $\alpha$.
d) $M=3$, $i=1,\dots,8$, e) $M=4$, $i=1,\dots,7$, f) $M=5$, $i=1,\dots,6$.}
  \label{Fig:avrAll2} 
\end{figure*}

The  functions $ P_{3,i}^{max} $, $i=1,\dots,8$, $ P_{4,i}^{max} $, $i=1,\dots,7$, and   $ P_{5,i}^{max}$, $i=1,\dots,6$, 
in terms of the mean values and root-mean-square deviations with respect to $\alpha$ are shown in Fig.\ref{Fig:avrAll}. 
The mean-values depicted in Fig.\ref{Fig:avrAll}a,b,c  show that the dependence on $\lambda$ becomes more significant for those clusters which include the 5th and 6th spins (i.e., $i\sim \frac{N}{2}$) when  the concurrence $C_{56}$ is involved into the definition (\ref{E21}). In addition the root-mean-square deviations (Fig.\ref{Fig:avrAll}d,e,f) show that the  effect of $\alpha$ increases with an increase in $\lambda$ (over the interval $\frac{1}{2}\le \lambda\le 1$),

The mean values and root-mean-square deviations of the geometric mean $P_{M,i}^{max}$ with respect to $\lambda$
defer from those  with respect to $\alpha$, see Fig.\ref{Fig:avrAll2}. In particular, not all the  mean values are monotonic functions of $\alpha$. Those of them which include $C_{56}$ in their definition have the maximum. Moreover, all the root-mean-square deviations are also non-monotonic functions. They have either  maximum (those that include $C_{56}$ in their definitions) or minimum (all others). Therefore, the effect of the eigenvalue parameter $\lambda$ depends mainly on the position of the cluster in the chain. 


%
In addition, all the geometric means $P_{M,i}$ involving $C_{56}$ in their definition (\ref{E21}) have the critical values  $\lambda_c$ and  $\alpha_c$ (both depending on $M$ and $i$)
such that the clusters exist in the intervals $\lambda_c(M,i)\le \lambda \le 1 $ and $0\le \alpha\le \alpha_c(M,i)$.
These critical values for $M=3,4,5$ are following.
\begin{eqnarray}
M=3:\;\;&&\lambda_c(3,4)= \lambda_c(3,5) = 0.82,\;\;\alpha_c(3,4)=\alpha_c(3,5) = 0.76,\\\nonumber
M=4:\;\;&&\lambda_c(4,3)= \lambda_c(4,4)= \lambda_c(4,5) = 0.82,\;\;\alpha_c(4,3)=\alpha_c(4,4) =\alpha_c(4,5) = 0.66,\\\nonumber
M=5:\;\;&&\lambda_c(5,2)= \lambda_c(5,5)=0.84,\;\;\;\  \lambda_c(5,3) =\lambda_c(5,4) = 0.82,\;\;\\\nonumber
&&\alpha_c(5,2)=\alpha_c(5,5) =0.64,\;\;\;\alpha_c(5,3) =\alpha_c(5,4) = 0.62.
\end{eqnarray}

The geometric mean $P_{M,i}$ characterizes the overlap of different pairwise concurrences in the cluster of $M$ particles. This function depends on the shape of each concurrence, their maximal values and instants of these maxima. In general, clusters $P_{M,i}$ become less entangled with an increase in $M$ and $i$, which is demonstrated in Figs.\ref{Fig:avrAll} and \ref{Fig:avrAll2}.

\subsection{Large clusters}

\begin{figure*}
\subfloat[]{\includegraphics[scale=0.3,angle=0]{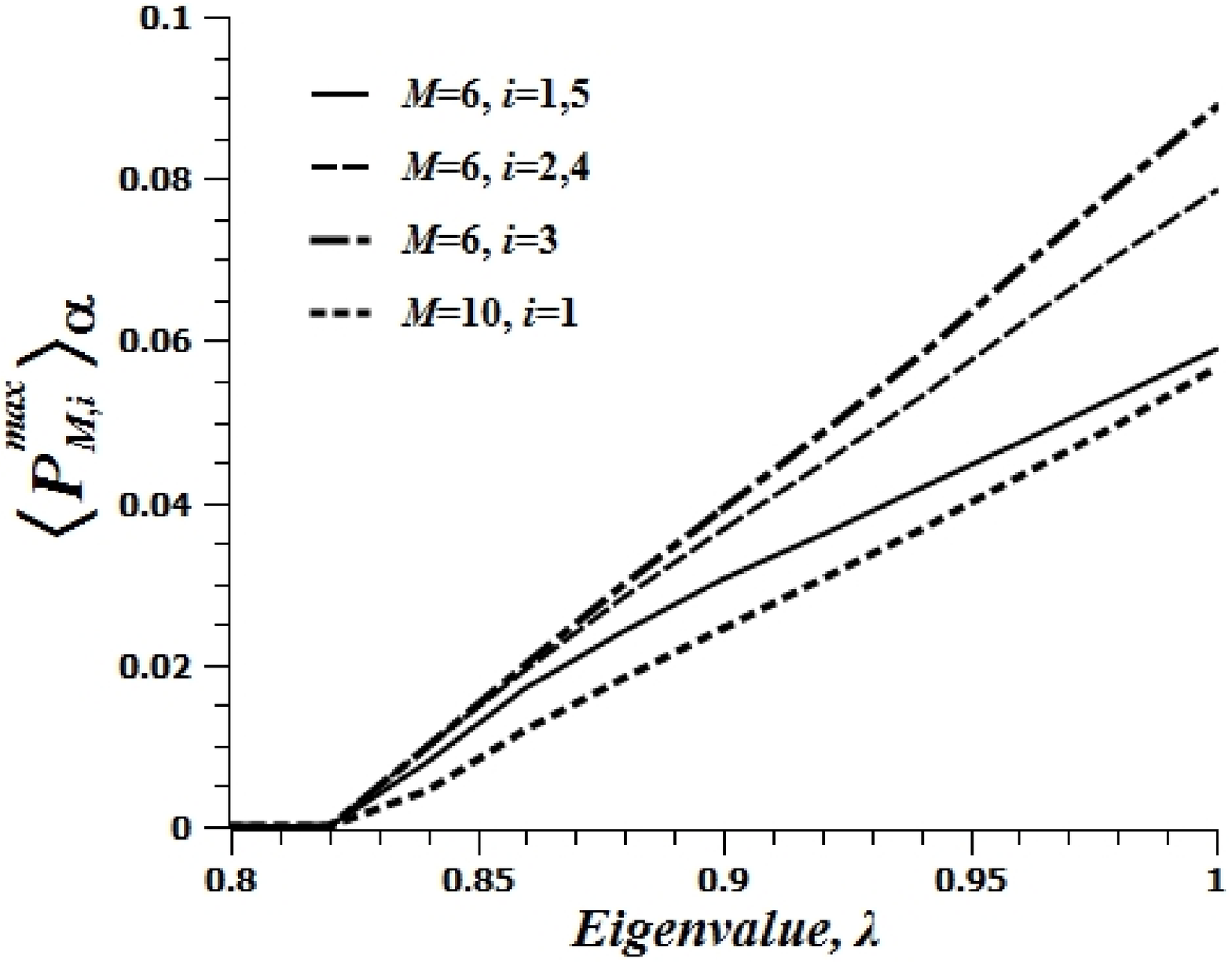
}}
\subfloat[]{\includegraphics[scale=0.3,angle=0]{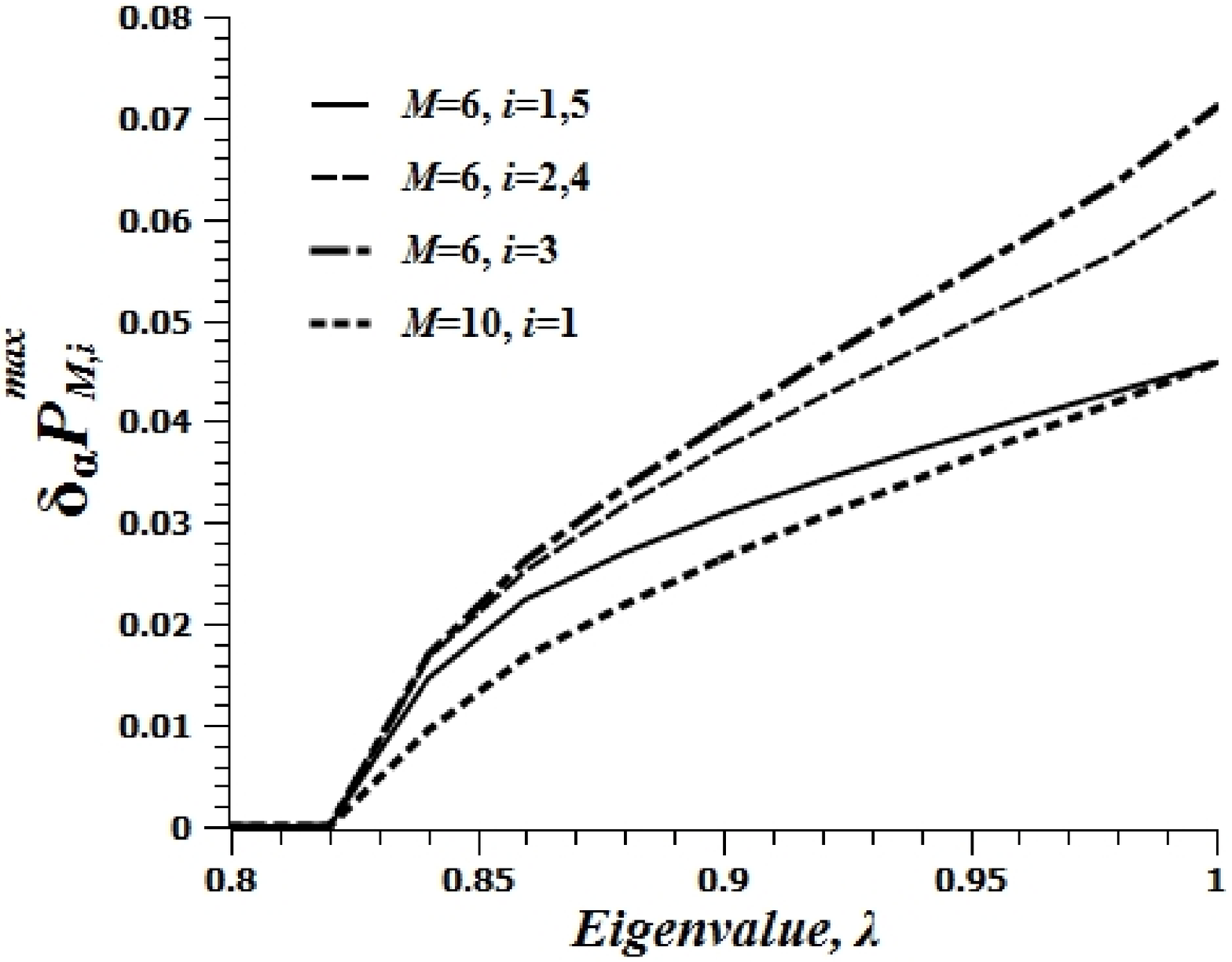
}}\\
\subfloat[]{\includegraphics[scale=0.3,angle=0]{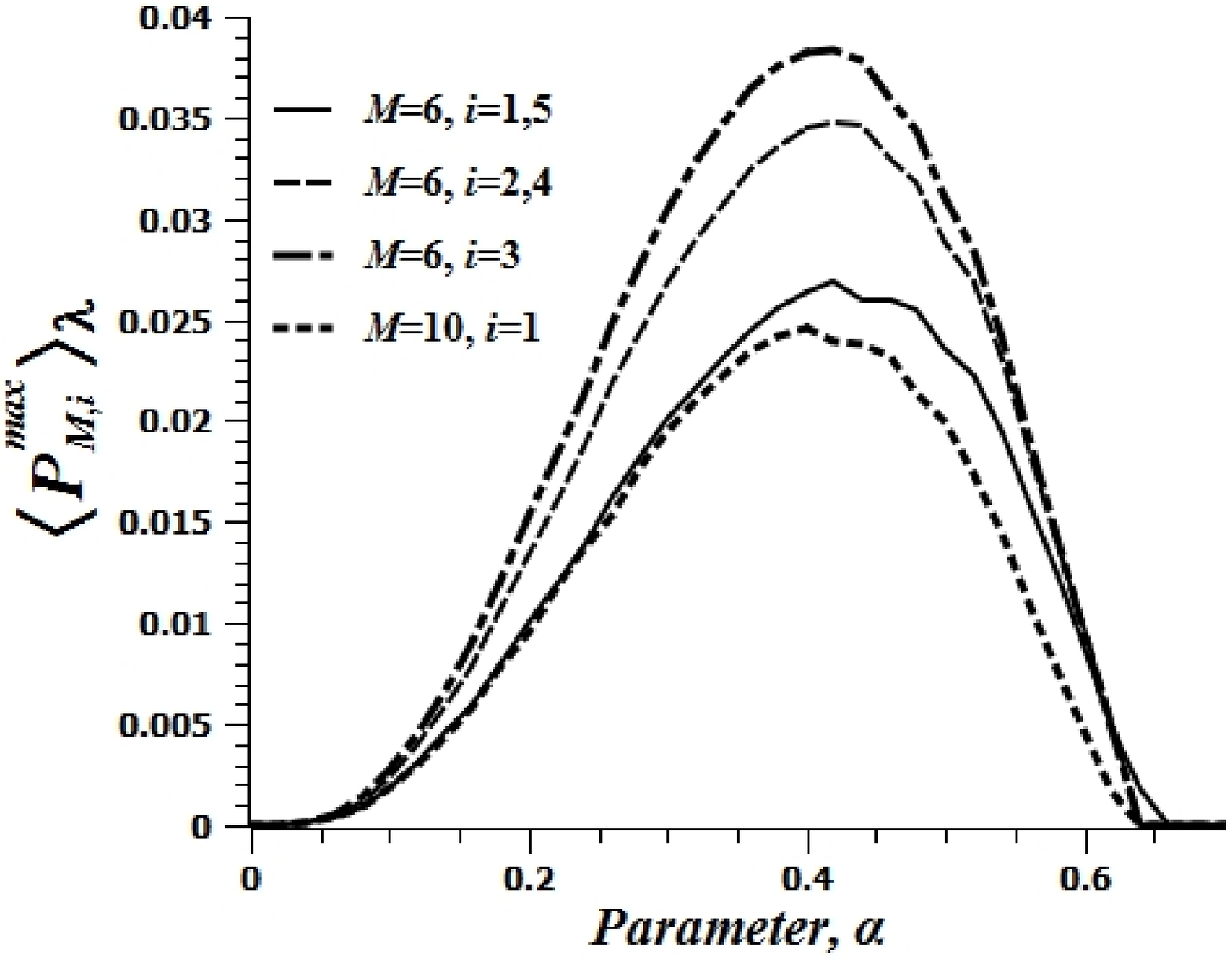
}}
\subfloat[]{\includegraphics[scale=0.3,angle=0]{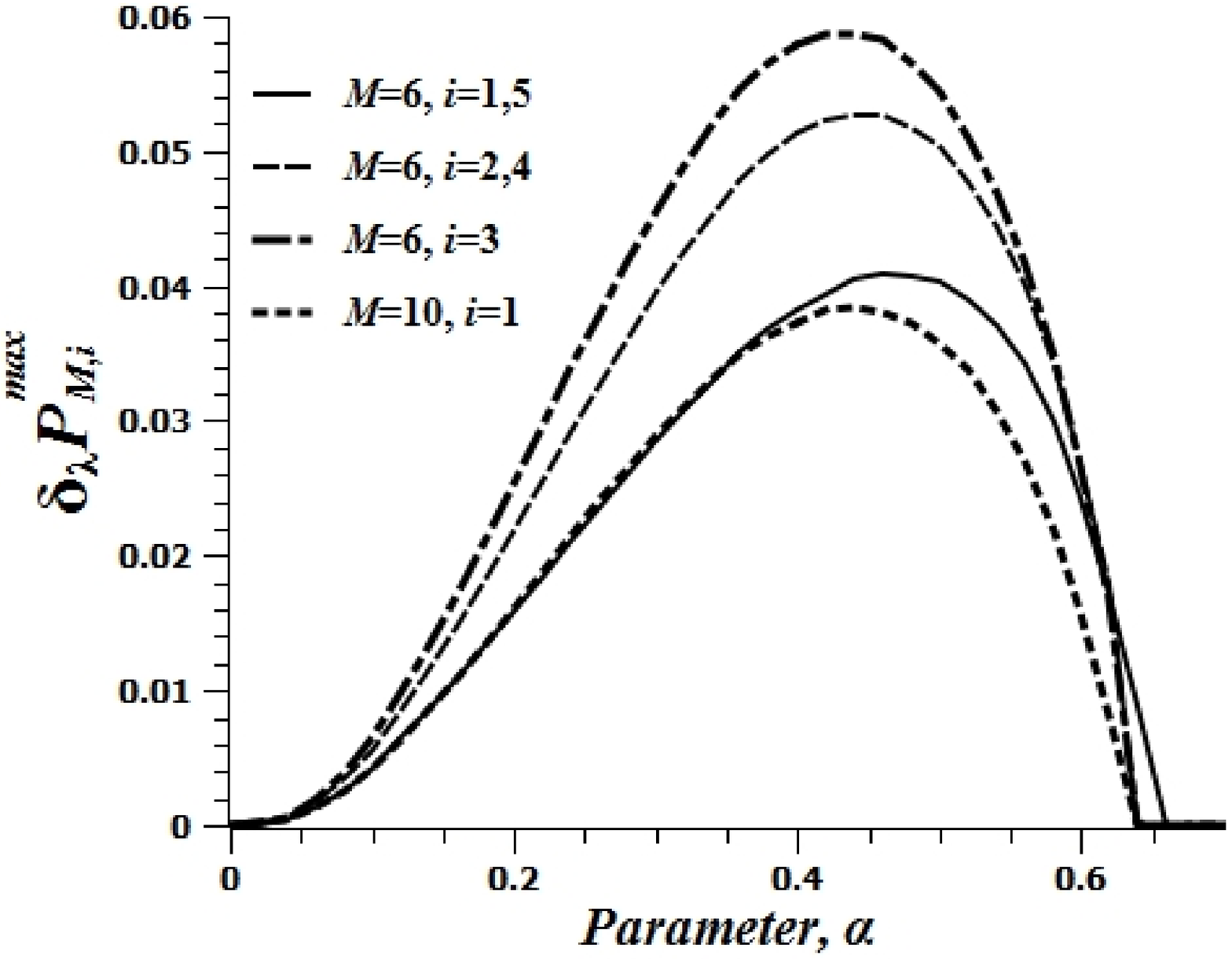
}}
\caption{The mean values and root-mean-square deviations for $M=6$ (solid lines) and $M=10$ (dashed lines).
a) The mean values  $\big\langle P_{M,i}^{max}\big\rangle_\alpha$ as  functions of $\lambda$;
b) The mean values  $\big\langle P_{M,i}^{max}\big\rangle_\lambda$ as  functions of $\alpha$;
c) The root-mean-square deviations  $\delta_\alpha P_{M,i}^{max}$ as  functions of $\lambda$;
d) The root-mean-square deviations  $\delta_\lambda P_{M,i}^{max}$ as  functions of $\lambda$}
  \label{Fig:avrAllOthers} 
\end{figure*}

For completeness, we also { discuss the formation of large clusters ($M>5$) and}  depicture  
the geometric mean $P_{M,i}^{max}$ for $M=6$ and $M=10$ in Fig.\ref{Fig:avrAllOthers} in terms of mean values and root-mean-square deviations. All these functions involve the concurrence $C_{56}$  in their definition (\ref{E21}), so that  
the critical values $\lambda_c$ and $\alpha_c$ exist for any $i$. For the fixed $M$, 
both mean values of  functions $P_{M,i}^{max}$ (i.e., with respect to $\alpha$ and $\lambda$) and both 
root-mean-square deviations get the maximal values {for}  $i$ corresponding to the middle of its variation interval ($i=3$ for $M=6$), as shown in Fig.\ref{Fig:avrAllOthers}. 

\subsection{Cluster's life-time}
\begin{figure*}
\subfloat[]{\includegraphics[scale=0.3,angle=0]{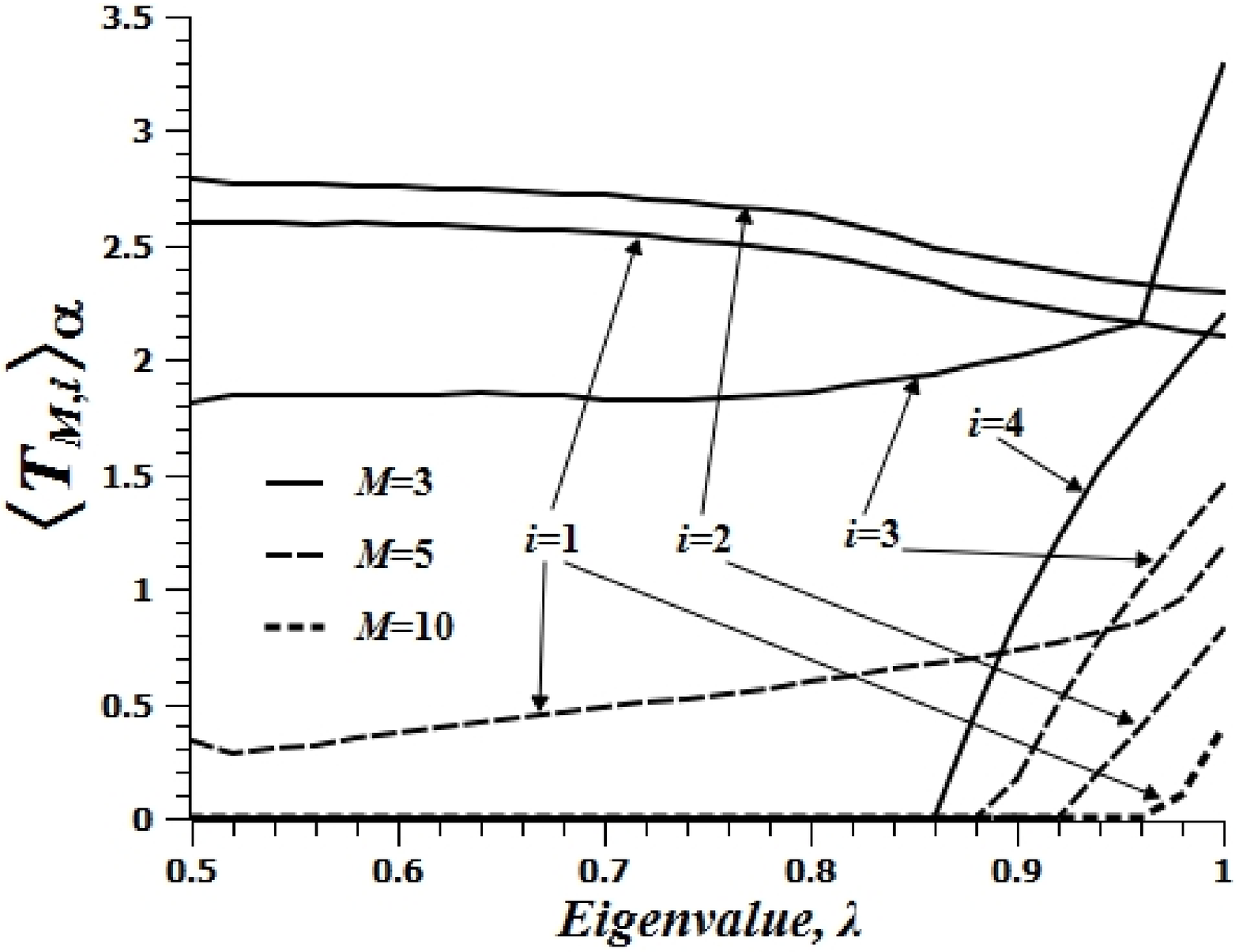
}}
\subfloat[]{\includegraphics[scale=0.3,angle=0]{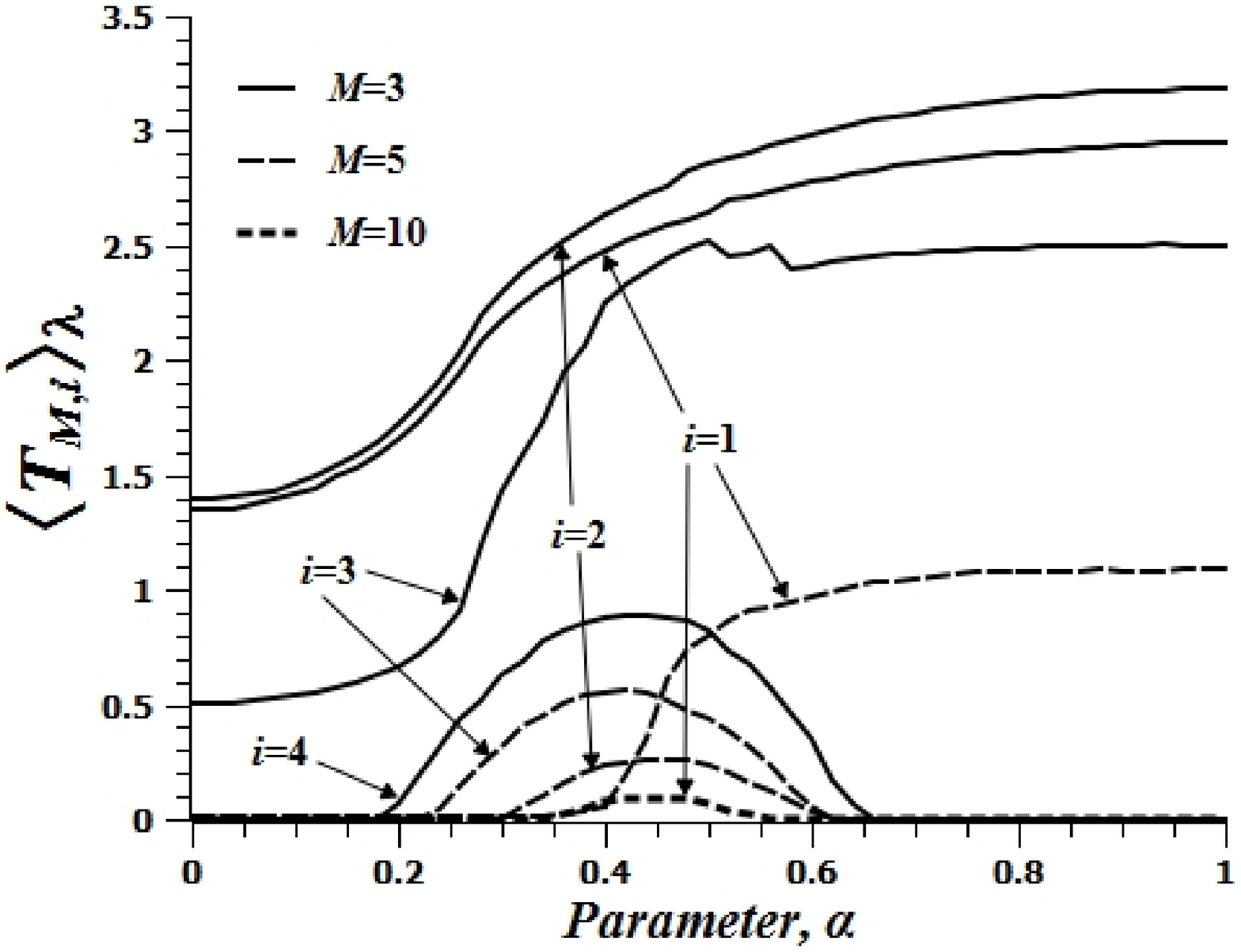
}}\\
\subfloat[]{\includegraphics[scale=0.3,angle=0]{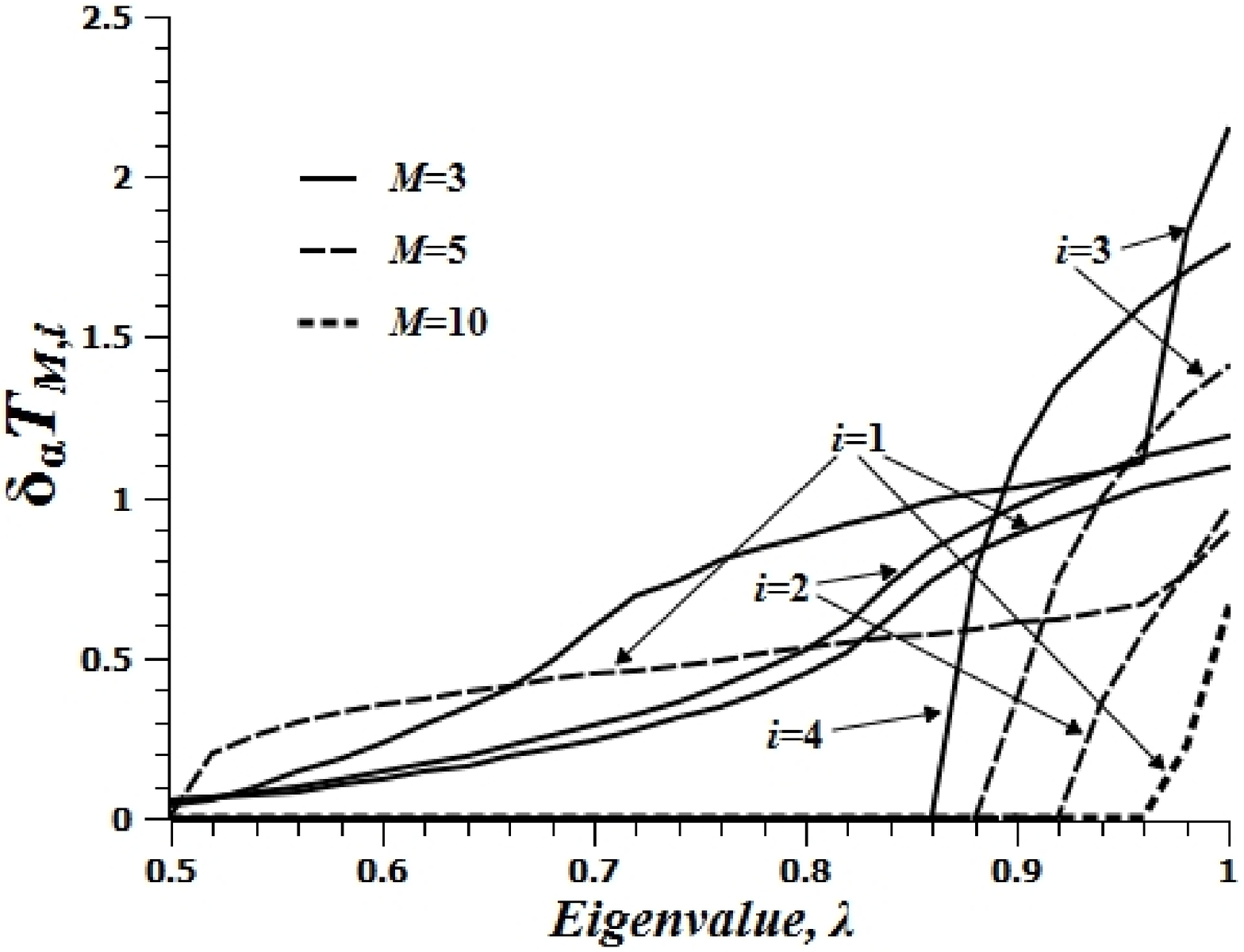
}}
\subfloat[]{\includegraphics[scale=0.3,angle=0]{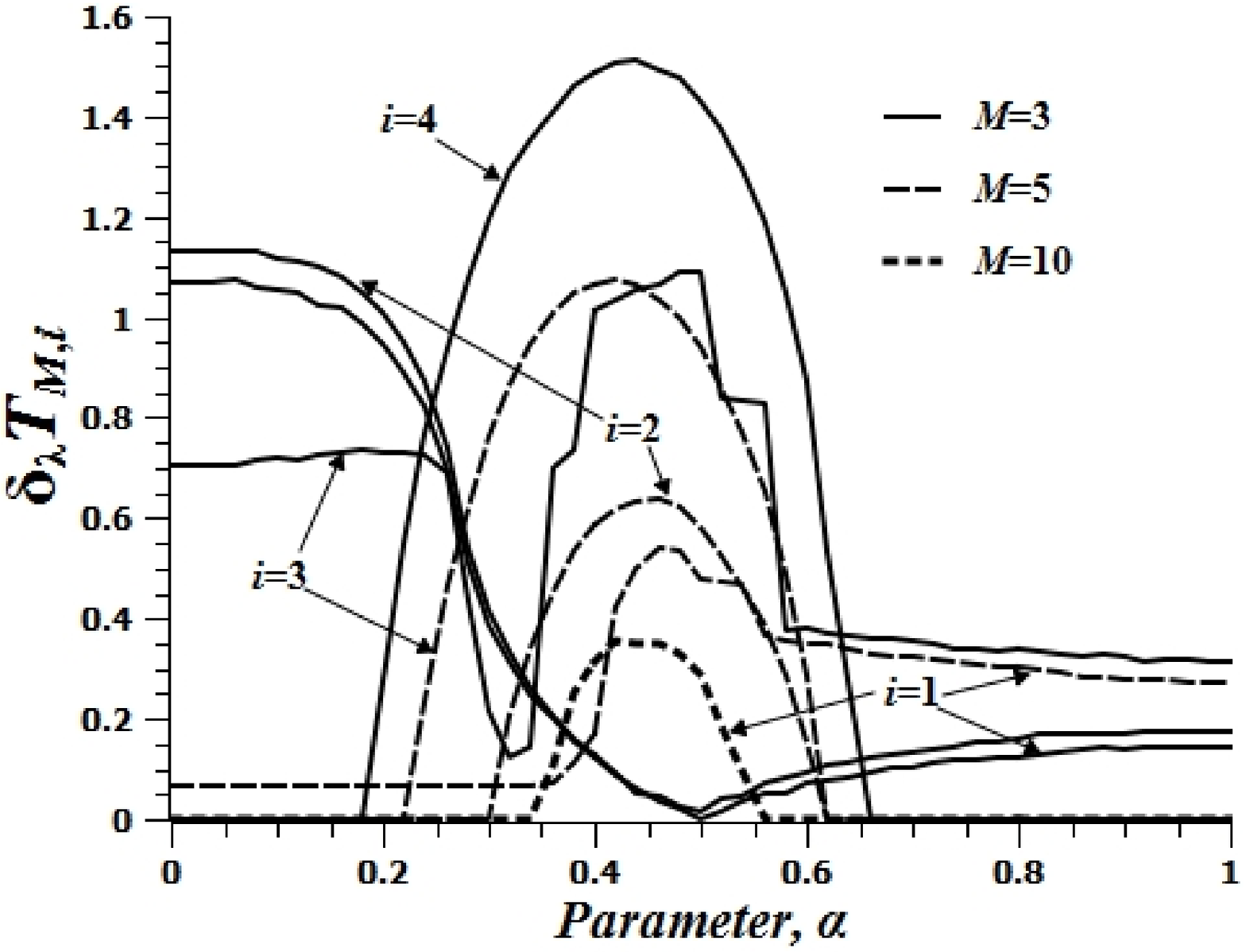
}}
\caption{The  life-times of clusters. a) The mean values $\big\langle T_{M,i}\big\rangle_\alpha$
as functions of $\lambda$;  b) The mean values $\big\langle T_{M,i}\big\rangle_\lambda$
as functions of $\alpha$; 
c) the root-mean-square deviations  $\delta_\alpha T_{M,i}$ as functions of $\lambda$;
d) the root-mean-square deviations  $\delta_\lambda T_{M,i}$ as functions of $\alpha$. }
  \label{Fig:width} 
\end{figure*}

Clearly, the clusters exist during some time intervals, which we call the cluster's  life-time.
We say that the cluster is destroyed if $P_{M,i}^{max}<\varepsilon_C$, where   $\varepsilon_C$ is some conventional value which is taken to be equal 0.1 in this paper. We estimate the life-time by the formula 
\begin{eqnarray}\label{T}
T_{M,i}= t_{r}-t_l,
\end{eqnarray}
where $t_{r}$ and $t_l$  are such  that 
$\varepsilon_C < P_{M,i} \le P_{M,i}^{max}$ over the interval $t_{r}\le t \le t_l$.  
This period depends on both $\lambda$ and $\alpha$. To illustrate this dependence, we depicture  the mean values ($\big\langle T_{M,i} \big\rangle_\alpha(\lambda)$ and 
$\big\langle T_{M,i} \big\rangle_\lambda(\alpha)$) and  root-mean-square deviations
 ($\delta_\alpha T_{M,i}(\lambda)$ and 
$\delta_\lambda T_{M,i} (\alpha)$) in Fig.\ref{Fig:width}. These figures show that the clusters,  whose geometric means $P_{M,i}$ involve $C_{56}$ in  their  definition (\ref{E21}), have restricted domain in the plane $(\lambda,\alpha)$.  
Fig.\ref{Fig:width}a demonstrates that  $\big\langle T_{M,i}\big\rangle_\alpha$ 
slightly depends on $\lambda$ for those clusters which do not involve $C_{56}$ in the definition of $P_{M,i}$. 
The dependence  of   $\big\langle T_{M,i}\big\rangle_\lambda$   on $\alpha$ 
is significant for $\alpha\lesssim 0.65$ which can be seen in
Fig.\ref{Fig:width}b. For $\alpha\gtrsim 0.65$, $\big\langle T_{M,i}\big\rangle_\lambda$ is almost constant.
Fig.\ref{Fig:width}c shows that the spread of $T_{M,i}$ at {the} fixed $\lambda$ and different  $\alpha$ becomes more significant with an increase in $\lambda$. Similarly, Fig.\ref{Fig:width}d shows that the spread of $T_{M,i}$ at a fixed $\alpha$ and different 
$\lambda$ is (generally)   significant for $\alpha \lesssim 0.65$, except the cases $M=3$, $i=1,2$, when this spread is valuable for  $\alpha \lesssim 0.4$.  Generally,  the life-time decreases with an increase in $M$. 

\section{Conclusions}
\label{Section:conclusion}
Although the correlations between the sender and receiver may vanish during the whole evolution period, other pairwise correlations arise during the evolution and supplement the information propagation.
While the entanglement between any two particles in an evolutionary spin chain  is located in time, the sum $S$ of all pairwise entanglements    { can be significant}  over the whole evolution period. One can say that the entanglement propagates from a given pair of entangled particles to another one; therefore we refer to this sum as the relay entanglement. {Thus, the relay entanglement (the assembly of pairwise quantum entanglements) }  supplements   information propagation.  It is { natural} to assume that the correlations between the nearest neighbors are  most important. However, this is not always true because the entanglements between the remote nodes prevail in certain cases
{ which is justified by the partial sums $S_m$, whose  $S_m^{max}$  and  $S_m^{min}$ (\ref{Sm}-\ref{min})  are described  in terms of their mean values  and    
  root-mean-square deviations with respect to $\alpha$ and $\lambda$, eqs.(\ref{mean}-\ref{dev2}), in the model of 10-node homogeneous spin-1/2 chain with the initial state  symmetrical with respect to the change $S\leftrightarrow R$.}

  During the evolution,  the clusters of entangled particles arise (i.e., the clusters where all pairwise entanglements are significant), where  geometric mean $P_{M,i}$ (\ref{E21}) serves 
  as the {characteristics} of entanglement. We study its dependence on the initial-state parameters $\lambda$ and $\alpha$ using the mean value and root-mean-square deviation with respect to the parameters $\alpha$ and $\lambda$ { and graphically show that  the geometric mean  $P_{M,i}$ can be  almost constant  over some  intervals of  these parameters, while it can significantly vary over other intervals.  For  the large clusters ($M>5$ in our case), in addition, there are intervals $\lambda <\lambda_c$ and $\alpha>\alpha_c$  where these clusters do not exist. }
  The life-time of the clusters  decreases with an increase in the size of a cluster, however, the cluster's position in the chain  also affects  the lifetime. { The large-size clusters are short-living in this model, while small-size
    clusters of such type can be candidates for  quantum registers. These registers carry information encoded into the initial state of a quantum system (for instance, into the sender). Therefore, applying a proper unitary transformation to the localized entangled  cluster, we can affect  its state  and consequently the state registered at the receiver. This phenomenon can be used for realizing certain quantum gates, where the selected cluster serves as a control element. }

{ This work is partially supported by the program of the Presidium of RAS No. 5 ''Electron resonance, spin-dependent electron effects and spin technology''  and by the Russian Foundation for Basic Research, grant No.15-07-07928.}

\end{document}